\documentclass[acmsmall,screen,nonacm]{acmart}
\usepackage{tikz}
\usepackage{multirow, listings}
\usepackage{pifont}
\usepackage{colortbl} 
\usepackage{subcaption}
\usepackage{amsmath}
\usepackage{pgfplots}
\usepackage{graphicx, subcaption}
\usepackage{tcolorbox}
\usepackage{soul}
\usepackage{wrapfig}
\usepackage{balance}
\usepackage{makecell}
\usepackage{algorithm}
\usepackage{algorithmicx}
\usepackage{algpseudocode}
\usepackage{microtype}
\usepackage{xcolor}
\usepackage{booktabs}
\usepackage{tabularx}
\usepackage{caption}
\usepackage{subcaption}
\usepackage{hyperref}
\usepgfplotslibrary{groupplots}
\usepackage{array}
\usepackage{ragged2e}
\usepackage{tabularx,makecell,booktabs,array}
\usetikzlibrary{arrows.meta,positioning,fit,calc}
\usepackage{enumitem}
\newif\ifarxiv

\arxivtrue
\usepackage{xspace}
\usepackage{listings}
\usepackage{xcolor}
\usepackage{booktabs}
\usepackage{pifont}
\usepackage{comment}

\newcommand{\tool}{\textsc{Holey}\xspace}

\definecolor{bindbg}{RGB}{255,232,160}
\newcommand{\bnd}[1]{{\setlength{\fboxsep}{1pt}\colorbox{bindbg}{\ensuremath{#1}}}}
\newcommand{\bndt}[1]{{\setlength{\fboxsep}{1pt}\colorbox{bindbg}{\strut #1}}}
\lstdefinestyle{pyspark}{
  language=Python,
  basicstyle=\ttfamily\small,
  columns=fullflexible,
  keepspaces=true,
  showstringspaces=false,
  breaklines=true,
  frame=single,
  numbers=left,
  numberstyle=\ttfamily\tiny,
  xleftmargin=1.2em,
  keywordstyle=\color{blue}\bfseries,
  commentstyle=\color{gray}\itshape,
  stringstyle=\color{teal},
  identifierstyle=\color{black},
  emph={spark,createDataFrame,groupBy,agg,alias,first,filter,limit,count,sum},
  emphstyle=\color{purple}
}
\usepackage{mathpartir}
\usepackage{stmaryrd}
\newcommand{\sem}[1]{\llbracket #1 \rrbracket}
\newcommand{\Qpre}{Q_{\mathrm{pre}}}
\newcommand{\Qpost}{Q_{\mathrm{post}}}
\newcommand{\eqbag}{=}
\newcommand{\vS}{\bar\Sigma}
\newcommand{\Bag}[1]{\mathrm{Bag}\langle #1 \rangle}
\newcommand{\tst}[2]{\langle #1 \rangle \rhd #2}
\newcommand{\rn}[1]{\textsc{#1}}

\newcommand*{\labelalias}[2]{%
  \AtBeginDocument{\expandafter\let\csname r@#1\expandafter\endcsname\csname r@#2\endcsname}}
\labelalias{A-app:rules}{app:rules}
\labelalias{A-app:p3}{app:p3}
\labelalias{M-fig:mp-rules}{fig:mp-rules}
\begin{document}

\title{Properties with Holes: Meta-Property-Based Testing of Data-Intensive Computing Frameworks}

\author{Yaoxuan Wu}
\affiliation{%
  \institution{University of California, Los Angeles}
  \city{Los Angeles}
  \country{USA}}
\email{thaddywu@cs.ucla.edu}

\author{Ingrid Lee}
\affiliation{%
  \institution{University of California, Los Angeles}
  \city{Los Angeles}
  \country{USA}}
\email{ingridlee@g.ucla.edu}

\author{Ahmad Humayun}
\affiliation{%
  \institution{Virginia Tech}
  \city{Blacksburg}
  \country{USA}}
\email{ahmad35@vt.edu}

\author{Muhammad Ali Gulzar}
\affiliation{%
  \institution{Virginia Tech}
  \city{Blacksburg}
  \country{USA}}
\email{gulzar@cs.vt.edu}

\author{Miryung Kim}
\affiliation{%
  \institution{University of California, Los Angeles}
  \city{Los Angeles}
  \country{USA}}
\email{miryung@cs.ucla.edu}

\begin{abstract}
Data-intensive computing frameworks such as Apache Spark and Flink keep expanding their APIs with new data types, expressions, and operators. Their most damaging bugs return silently wrong results rather than crashing, which tools that detect only crashes or exceptions cannot see. Property-based testing can catch such bugs, but only if someone writes a property with a strong oracle, and writing properties one API at a time cannot keep pace with this growth. We observe that many such properties are instances of a few textbook principles: congruence of equivalent expressions, relational-algebra equivalences, and aggregate decomposition. Each principle holds not only for an isolated operator but also for any workload built around it. 

We present \tool, which introduces \emph{meta-properties}: properties with holes. A meta-property fixes how a workload is transformed and which relation its outputs must satisfy, but leaves open the expressions or operators being related. A developer fills the holes with a short binding. \tool then turns the resulting concrete property into many executable tests by generating schema-valid multi-stage workloads and input data around it. Each meta-property has a typing rule over its holes. A binding is checked once against this rule, and we prove that every test generated from a type-correct binding is schema-valid. Seven meta-properties yield 207 concrete properties, with bindings averaging under 20 lines. \tool found 17 previously unknown wrong-result bugs in the latest releases of Spark and Flink, 9 of which need relations that existing SQL testing oracles such as TLP and NoREC do not state, and its meta-properties express 55 of 65 testable historical Spark correctness issues. Its generated workloads cover 1.9$\times$ the branches covered by CometFuzz.

\end{abstract}

\maketitle

\section{Introduction}
Property-based testing (PBT) checks a system against a given property over
many generated inputs, but each property is still written by hand, one API at
a time, and this cost grows with the API under test. The problem is acute for
data-intensive computing frameworks such as Apache Spark~\cite{zaharia2010spark}
and Apache Flink~\cite{flink_vldb2015}, which form the backbone of modern data
analytics and whose APIs keep growing. As
Figure~\ref{fig:spark-expression-api-growth} shows, Spark's built-in
expressions alone grew from 267 in version~2.3 to 485 in version~4.0, alongside
new DataFrame operators and nested types such as arrays and structs. Checking
each API once is not enough: the query optimizer rewrites the plan around it,
so it must be checked in the context of the workloads that use it. For
example, Spark's \texttt{ColumnPruning} rule, which removes unused columns,
handles a single \texttt{explode}, which flattens an array into rows,
correctly, but returns wrong results when \texttt{explode} is chained across
nested structs and followed by a column drop~\cite{SPARK-39854}. The effort of
writing properties thus grows with both the number of APIs and the number of
contexts in which each must hold, and hand-written properties cannot keep pace.

\ifarxiv
\begin{figure}[t]
    \centering
    \includegraphics[width=0.55\linewidth]{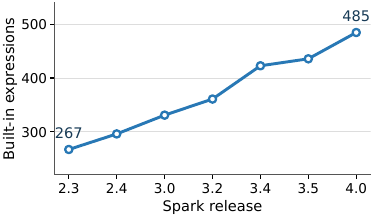}
    \caption{Spark's built-in expression APIs grew from 267 (v2.3) to 485 (v4.0).
Writing one property per API cannot keep pace; a meta-property covers
many APIs, each with a short binding.}
    \label{fig:spark-expression-api-growth}
\end{figure}
\fi
These frameworks are prone to logic bugs that silently return wrong results
rather than crashing~\cite{spark_jira_spark33726,spark_jira_spark49000}.
Fuzzers that detect crashes or uncaught exceptions cannot see such bugs. PBT
can catch them, because its oracle checks a semantic property of the output.
Database testing has shown how effective such oracles
are~\cite{rigger2020finding_tlp,norec_fse2020,pqs_osdi2020}.
TLP~\cite{rigger2020finding_tlp}, for example, checks that partitioning a query
by a predicate preserves its result. These oracles are built for SQL, so they
state no relations for constructs that DataFrame workloads use routinely, such
as user-defined functions written in Python and higher-order functions over
nested arrays. Moreover, each oracle is a separate test procedure that
implements one relation by hand. Covering a new function or operator then
means building another oracle, which brings back the per-API cost of
hand-written properties.

We present \tool, which makes properties reusable. Many properties one would
write for these frameworks are instances of a few principles that
textbooks state and query optimizers rely on: congruence of equivalent
expressions~\cite{harper2016pfpl}, relational-algebra
equivalences~\cite{silberschatz2019dbconcepts}, and aggregate
decomposition~\cite{gray1997datacube}. The principles are not new. \tool{} turns each of them
into a \emph{meta-property}, a property with holes, and checks it inside many
generated multi-stage workloads. A meta-property is a
reusable skeleton that fixes how a workload is transformed and which relation
its outputs must satisfy, leaving open the expressions or operators being
related. A developer fills the holes with a short binding. For example,
binding aggregation decomposition to \texttt{avg}, with the relation that \emph{the
global average lies between the smallest and largest per-group averages},
yields a concrete property.
Everything else is reusable: \tool's generators, shared by Spark and Flink,
build the surrounding workload, schemas, and data, placing the property
wherever its types allow in a multi-stage workload. A single short binding
thus yields many tests across generated workloads and input data, each with a
strong oracle derived from the property, where developers would otherwise
write each test and its expected output by hand.

A binding states what its holes must satisfy. For instance, the congruence binding
$e_1 \mapsto$~\texttt{forall(arr, p)} and
$e_2 \mapsto$~\texttt{not exists(arr,\allowbreak{} lambda x:\allowbreak{} not p(x))}
states $e_1 \equiv e_2$. The holes are typed, so a binding is checked once
against the typing rule of its meta-property, much as a compiler checks a
function body against the types of its parameters. Every test generated from a
type-correct binding is then schema-valid: each column that it references
exists with the expected type. We implement the check with the engines' own
type inference, and all our bindings pass it.

Realizing a concrete property as an executable PBT requires embedding it
within a schema-valid DataFrame program. This is hard because
a binding constrains where it may appear: its holes need columns of particular
types, and some expressions are legal only at certain positions, such as an
aggregate only in an aggregation list, so the workload must be built around the
binding. \tool{} builds the workload around the binding in a single pass that is
schema-aware, dependency-ordered, and binding-guided
(Section~\ref{sec:realization}), so that each concrete property is exercised across diverse schemas,
workloads, and input data.

We evaluate \tool{} on Apache Spark~4.2.0 and Apache Flink~2.3.0. It found 17
previously unknown wrong-result bugs, 7 in Spark and 10 in Flink. Nine of them
need relations that existing SQL testing oracles such as TLP and NoREC do not
state, and neither
SQLancer~\cite{sqlancer_repo} nor CometFuzz~\cite{cometfuzz_github} could have
found any of them. The approach generalizes across the two frameworks: both
share one workload and data generator, and each engine adds only an adapter
and per-meta-property templates. The meta-properties are also expressive: of
the 65 testable in-scope Spark correctness issues filed between 2020 and 2025,
they express 55.
Finally, generation scales: Spark accepts every generated workload without an
analysis error as schemas grow to 80 columns and workloads to
depth 96, and these workloads cover 1.9$\times$ the branches that CometFuzz
covers.
Our contributions are as follows.
\begin{enumerate}[leftmargin=*,itemsep=2pt,topsep=2pt,parsep=0pt,partopsep=0pt]
\item \textbf{Meta-properties.}
We introduce meta-properties, properties with holes that turn three textbook
principles (congruence of equivalent expressions, relational-algebra equivalences,
and aggregate decomposition) into laws over any multi-stage workload. Seven meta-properties yield 207 concrete properties
with bindings averaging under 20 lines, including relations that existing SQL testing
oracles do not state, such as UDF-to-builtin equivalences and higher-order
expression rewrites. 

\item \textbf{Typing rules.}
We give each meta-property a typing rule that divides its premises between the
binding and the generator, and prove that a binding checked once against its
rule yields only schema-valid tests. We implement the check with the type
inference of Spark and Flink.

\item \textbf{Realization.}
We design a realization algorithm that builds schema-valid multi-stage
workloads and input data around a binding in a single schema-aware,
dependency-ordered, and binding-guided pass. Its workload and data
generator is shared by our implementations for Apache Spark and Apache Flink.

\item \textbf{Evaluation.}
\tool{} found 17 previously unknown wrong-result bugs in Spark and Flink, all
reported to the developers. Its meta-properties express 55 of 65 testable
in-scope Spark correctness issues filed from 2020 to 2025, and Spark accepts its generated
workloads without an analysis error up to 80 columns and depth 96. They cover
1.9$\times$ the branches of CometFuzz. 
\end{enumerate}

Property-based testing automates input generation but leaves open where
properties come from: each one is still conceived and written by hand. For
data-intensive computing frameworks, \tool{} answers this question.
Properties come from principles that the frameworks already obey, and a developer
writes each principle once, as a meta-property, and reuses it many times: across
APIs through short bindings, across workloads and input data through
generation, and across engine versions as the API grows. Strong oracles are
also what automated test generation lacks: LLM-based test generators can
produce inputs and test code, but remain bottlenecked by the difficulty of
specifying strong oracles (i.e., expected outcomes)~\cite{zhou2026change}. \tool{} addresses this limitation directly by specifying meta-properties that can be instantiated to serve as strong oracles. We expect this \emph{write once, reuse many times}
recipe to extend to other domains with rich algebraic structure, such as
tensor computation frameworks.

\begin{figure}[t]
\centering
\small
\renewcommand{\arraystretch}{1.15}
\setlength{\fboxsep}{5pt}
\fbox{\begin{minipage}{0.96\linewidth}
\textbf{\textcircled{\scriptsize 1} Meta-property P4 (Aggregation Decomposition).} For every workload $Q$, input, value column $c$, and key $k$:
\[
R\bigl(\ \sem{[\,Q;\ \mathsf{agg}(f(c))\,]},\ \ \sem{[\,Q;\ \mathsf{groupBy}(k).\mathsf{agg}(f(c))\,]}\ \bigr)
\]
\textbf{\textcircled{\scriptsize 2} Binding,} written once by a developer:\quad
$f = \bnd{\mathsf{avg}}$, \quad $R(G, L) = \bnd{\min L \le G \le \max L}$

\vspace{1ex}
\textbf{\textcircled{\scriptsize 3} Generated by \tool{}}, one of many tests:

\vspace{0.3ex}
\begin{tabular}{@{\hspace{1em}}l@{\hspace{1.5em}}l@{}}
\texttt{a = table(x: int) [INT\_MAX, INT\_MAX, 1]} & {\footnotesize\color{gray} generated data, with boundary values} \\
\texttt{b = table(y: int) [1, 2]} & \\
\texttt{Q = a.crossJoin(b)} & {\footnotesize\color{gray} upstream workload} \\
$c = \texttt{x}$,\ \ $k = \texttt{x}$ & {\footnotesize\color{gray} columns typed for \textsf{avg} and \textsf{groupBy}} \\
\end{tabular}

\vspace{1ex}
\textbf{\textcircled{\scriptsize 4} Check} on Flink 2.3.0:

\vspace{0.3ex}
\begin{tabular}{@{\hspace{1em}}l@{\hspace{1.5em}}l@{}}
$G = \sem{[\,\texttt{Q; agg(avg(x))}\,]}$ & $= 0$ \quad {\footnotesize\color{gray} (exact average 1431655765)} \\
$L = \sem{[\,\texttt{Q; groupBy(x).agg(avg(x))}\,]}$ & $= \{1,\ $\texttt{INT\_MAX}$\}$ \\
$\min L \le G \le \max L$ & {\color{red}violated}, Flink-1 in Table~\ref{tab:new-bugs} \\
\end{tabular}
\end{minipage}}
\caption{
Instantiating a meta-property with a binding. P4 has two holes, the
aggregate $f$ and the relation $R$. A developer writes only the highlighted
binding, and \tool{} generates everything else. The test shown, simplified from
the one that found Flink-1, exposes that bug.
}

\label{fig:motivating-instantiation}
\vspace{-2ex}
\end{figure}

\section{Background and Motivating Example}
\subsection{Background on DISC Workloads}
\label{sec:background}
Data-intensive scalable computing (DISC) engines such as Spark and Flink let developers write computations as workloads over \emph{DataFrames}, which Flink calls tables. A DataFrame is a typed tabular collection whose schema gives the names and types of its columns. A workload is a sequence of statements, each of which derives a new DataFrame from earlier ones, for example \texttt{df2 = df0.join(df1, "id")} followed by \texttt{df3 = df2.filter(F.col("tag") != "z")}. These dependencies form a directed acyclic graph whose sink is the result of the workload. Transformations are evaluated lazily: each one only extends a logical plan, which the engine optimizes and executes when the result is requested. Besides relational operators such as join, filter, and aggregation, workloads use user-defined functions, higher-order functions over arrays, and nested data.

\paragraph{Notation.} We write $[t_1; \dots; t_n]$ for the workload of statements $t_1, \dots, t_n$ and $[Q_1; Q_2]$ for the workload that runs $Q_2$ on the output of $Q_1$. We write $\sem{Q}D$ for the output of workload $Q$ on source data $D$, a bag of rows, and we omit $D$ when it is clear from the context, in particular in properties, which hold for every $D$.

\subsection{Motivating Example}
\label{sec:motivating}
\tool{} tests a DISC engine against properties that hold for every workload and every input. Consider the meta-property P4 (Aggregation Decomposition) in part~1 of Figure~\ref{fig:motivating-instantiation}: for any workload $Q$, an aggregate $f$ over all rows of $Q$'s output is related by a relation $R$ to the same aggregate computed per group. The aggregate $f$ and the relation $R$ are holes, which a developer fills with a \emph{binding} (part~2), here $f := \mathsf{avg}$ and $R(G, L) := \min L \le G \le \max L$. $R$ is a bound rather than an equality because an average cannot be recomputed from per-group averages alone~\cite{gray1997datacube}. The result is a concrete property: on any data and any deterministic upstream workload, the global average of a column must lie between its smallest and largest per-group averages.

\tool{} fills everything else (part~3): the source tables with their schemas and data, such as a table \texttt{a} whose integer column \texttt{x} holds \texttt{INT\_MAX} twice and 1 and a two-row table \texttt{b}, the upstream workload $Q$, such as the cross join of \texttt{a} and \texttt{b}, and the columns that \texttt{avg} and \texttt{groupBy} apply to, such as \texttt{x} for both. For readability, we show workloads as DataFrame pseudocode or SQL throughout the paper, while \tool{} emits PySpark and PyFlink programs.

On Apache Flink 2.3.0, this test fails (part~4): the global average is 0, whereas the per-group averages are 1 and \texttt{INT\_MAX}. This is Flink-1 in Table~\ref{tab:new-bugs}, a previously unknown bug that \tool{} found. The planner rewrites the average into a sum and a count, pushes them below the join, and computes the sum in the \texttt{INT} type, where it overflows.

A test of \texttt{avg} on its own would miss this bug, because Flink returns the correct average 1431655765 over \texttt{a} alone. Here, the cross join, which duplicates every row and thus leaves the average unchanged, triggers the faulty rewrite, so a hand-written test would have to anticipate both the join and the extreme values. \tool{} found the bug because a single short binding turns into many tests, each of which checks the same property in a different generated workload with different data.

\section{Meta-Properties}
\label{sec:mp}

\providecommand{\mphead}[2]{\multicolumn{2}{@{}p{\textwidth}@{}}{\textbf{#1}\ \ \emph{#2}}}
\begin{table}[t]
\caption{Seven meta-properties turn textbook principles into laws over any workload, and a short binding instantiates each of them. Highlighted symbols are the holes that a binding fills, with one example binding on the right. \tool{} generates the rest, including $Q$, $Q_{\mathrm{pre}}$, $Q_{\mathrm{post}}$, and the input. Read P1 as: if $e_1$ and $e_2$ are equivalent, then using either one at the same point of any workload gives the same output.}
\label{tab:mp-overview}
\centering
\small
\setlength{\tabcolsep}{4pt}
\renewcommand{\arraystretch}{1.3}
\begin{tabularx}{\textwidth}{@{}>{\raggedright\arraybackslash}X >{\raggedright\arraybackslash}p{4.4cm}@{}}
\toprule
\textbf{Law} & \textbf{Example binding} \\
\midrule
\mphead{P1: Higher-Order Expression Equivalence Rewrite.}{Replacing a higher-order expression with an equivalent one preserves the output of the whole workload.} \\
$\bnd{e_1} \equiv \bnd{e_2} \ \Rightarrow\ \sem{[\,Q_{\mathrm{pre}};\ \mathrm{op}(\bnd{e_1});\ Q_{\mathrm{post}}\,]} = \sem{[\,Q_{\mathrm{pre}};\ \mathrm{op}(\bnd{e_2});\ Q_{\mathrm{post}}\,]}$
 & $e_1 := \bnd{\mathsf{forall}(a, p)}$ \newline $e_2 := \bnd{\neg\,\mathsf{exists}(a, \lambda x.\,\neg p(x))}$ \\
\midrule
\mphead{P2: UDF Counterpart Equivalence Rewrite.}{Replacing a Python UDF with an equivalent builtin preserves the output of the whole workload.} \\
$\bnd{f} \equiv \bnd{e} \ \Rightarrow\ \sem{[\,Q_{\mathrm{pre}};\ \mathrm{op}(\mathsf{udf}(\bnd{f}));\ Q_{\mathrm{post}}\,]} = \sem{[\,Q_{\mathrm{pre}};\ \mathrm{op}(\bnd{e});\ Q_{\mathrm{post}}\,]}$
 & $f := \bnd{\lambda s.\ s[1\!:\!4]}$ (Python) \newline $e := \bnd{\mathsf{substring}(s, 2, 3)}$ \\
\midrule
\mphead{P3: Predicate Materialization with Downstream Adaptation.}{Filtering early gives the same output as tagging rows and filtering at the end.} \\
$\sem{[\,Q_{\mathrm{pre}};\ \mathsf{filter}(p);\ Q_{\mathrm{post}}\,]} = \sem{[\,Q_{\mathrm{pre}};\ \mathsf{withColumn}(k, p);\ \lceil Q_{\mathrm{post}}\rceil_k;\ \mathsf{filter}(k);\ \mathsf{drop}(k)\,]}$
 & none, $p$ is generated \\
\midrule
\mphead{P4: Aggregation Decomposition.}{An aggregate over all rows is related to the same aggregate per group.} \\
$\bnd{R}(\bnd{f}(B),\, \{\bnd{f}(B_i)\}) \ \Rightarrow\ \bnd{R}(\sem{[\,Q;\ \mathsf{agg}(\bnd{f})\,]},\ \sem{[\,Q;\ \mathsf{groupBy}(k).\mathsf{agg}(\bnd{f})\,]})$
 & $f := \bnd{\mathsf{count}}$ \newline $R(G, L) := \bnd{G = \textstyle\sum L}$ \\
\midrule
\mphead{P5: Predicate-Partition Decomposition.}{The rows on which a predicate is true, false, or null make up the whole output.} \\
$\sem{Q} = \sem{[\,Q;\ \mathsf{filter}(p)\,]} \uplus \sem{[\,Q;\ \mathsf{filter}(\neg p)\,]} \uplus \sem{[\,Q;\ \mathsf{filter}(p\ \mathsf{is\ null})\,]}$
 & none, $p$ is generated \\
\midrule
\mphead{P6: Intermediate Materialization Equivalence.}{Materializing an intermediate result preserves the output of the whole workload.} \\
$\sem{\bnd{M}} = \mathrm{id} \ \Rightarrow\ \sem{[\,Q_{\mathrm{pre}};\ Q_{\mathrm{post}}\,]} = \sem{[\,Q_{\mathrm{pre}};\ \bnd{M};\ Q_{\mathrm{post}}\,]}$
 & $M := \bnd{\mathsf{checkpoint}}$ \\
\midrule
\mphead{P7: Operator Input-Output Relations.}{The output of an operator stands in a fixed relation to its input.} \\
$\bnd{R}(B, \sem{\bnd{\mathrm{op}}}B) \ \Rightarrow\ \bnd{R}(\sem{Q},\ \sem{[\,Q;\ \bnd{\mathrm{op}}\,]})$
 & $\mathrm{op} := \bnd{\mathsf{filter}(p)}$ \newline $R(B, B') := \bnd{B' \subseteq B}$ \\
\bottomrule
\end{tabularx}
\par\smallskip\raggedright\footnotesize \emph{Notation.} $\sem{W}$ is the output of workload $W$, $=$ compares outputs as bags, $\mathrm{op}(e)$ is any operator that takes the expression $e$, and $B$ is any bag of rows.
\end{table}

A \emph{meta-property} is a semantic law with holes. It fixes how several related workloads are built and which relation their outputs must satisfy, but it leaves some parts open, such as the expressions being compared or the aggregate being decomposed. A \emph{binding} fills the holes on which the law depends, for example with a pair of equivalent expressions or with an aggregate and a relation. A meta-property together with a binding is a \emph{concrete property}, such as the property in Figure~\ref{fig:motivating-instantiation} that \emph{the average of a column over all rows lies between the smallest and the largest per-group average}. \tool{} fills the remaining holes, the workloads and their inputs, anew for each test (Section~\ref{sec:realization}). Because a concrete property relates the outputs of several workloads, it is a test oracle that needs no expected output.

Table~\ref{tab:mp-overview} lists the seven meta-properties. P1, P2, and P6 build on the congruence of semantic equivalence~\cite{harper2016pfpl}, P3, P5, and P7 on relational-algebra laws~\cite{silberschatz2019dbconcepts}, and P4 on the decomposition of aggregates~\cite{gray1997datacube}. None of these principles is new. Textbooks state them for an idealized semantics, and query optimizers assume that the engine obeys them. \tool{} checks whether it does: it turns each principle into a meta-property whose holes a short binding fills, and it checks the result inside many generated multi-stage workloads. We first explain each meta-property (Section~\ref{sec:mp-catalog}), then discuss when their oracles hold (Section~\ref{sec:mp-laws}), and finally show that a type-correct binding yields schema-valid tests (Section~\ref{sec:mp-typing}).

\subsection{The Seven Meta-Properties}
\label{sec:mp-catalog}
We group the meta-properties by the textbook principle that they build on.

\paragraph{Congruence: P1, P2, and P6.}
If two parts of a program compute the same function, replacing one with the other cannot change the program's result~\cite{harper2016pfpl}. DISC engines often offer several ways to compute the same thing, and each way runs through different code. These meta-properties use congruence to check that such alternatives agree, both on their own and inside any surrounding workload.

\textbf{P1: Higher-Order Expression Equivalence Rewrite.} Replacing a higher-order expression $e_1$ with an equivalent expression $e_2$ at any point of a workload leaves the output unchanged. For example, $\mathsf{forall}(a, p)$ is true when the predicate $p$ holds for every element of the array $a$, and $\neg\,\mathsf{exists}(a, \lambda x.\,\neg p(x))$ is true when no element violates $p$, so a binding can pair the two. \tool{} generates the array $a$ and the predicate $p$. Higher-order functions such as \texttt{forall}, \texttt{exists}, \texttt{transform}, and \texttt{aggregate} apply a lambda to the elements of an array, and engines implement each of them separately. P1 tests that these implementations agree, including on null elements and empty arrays, and that optimizer rewrites inside lambdas preserve their meaning.

\textbf{P2: UDF Counterpart Equivalence Rewrite.} Replacing a Python UDF $f$ with a builtin expression $e$ that computes the same function leaves the output unchanged. For example, the Python function $\lambda s.\ s[1\!:\!4]$ returns the second to fourth characters of a string $s$, and so does the builtin $\mathsf{substring}(s, 2, 3)$, which takes three characters from position 2 because positions start at 1.\footnote{For simplicity, the example ignores a null $s$, on which the Python function would fail. The registered binding returns null for a null input and samples the start position and the length.} The builtin runs inside the engine's JVM. The UDF runs in a separate Python process: the engine serializes each input value, sends it to Python row by row or in Arrow batches, and converts the result back to the type that the UDF declares. P2 tests that values keep their meaning across this boundary, including null, \texttt{NaN}, strings, and timestamps.

\textbf{P6: Intermediate Materialization Equivalence.} Inserting a materialization operator $M$, which returns its input unchanged, at any point of a workload leaves the output unchanged. For example, $\mathsf{checkpoint}$ computes an intermediate DataFrame, stores it, and continues from the stored copy. Other materialization operators cache the data at some storage level, or write it to Parquet or ORC and read it back. Because DataFrame programs are evaluated lazily, $M$ also changes what the optimizer sees: without $M$ it optimizes the whole plan, and with $M$ it optimizes the two parts separately. P6 tests that optimizer rewrites across the cut, and the storage formats themselves, preserve the data.

\paragraph{Relational-algebra laws: P3, P5, and P7.}
Relational algebra comes with laws about how operators combine, for example when a selection can move past another operator, and about what an operator's output contains~\cite{silberschatz2019dbconcepts}. Query optimizers apply these laws to rewrite plans. These meta-properties check that the rewritten plans still obey the laws inside generated workloads.

\textbf{P3: Predicate Materialization with Downstream Adaptation.} Filtering by a predicate $p$ early gives the same output as recording $p$ in a column and filtering at the end. For example, filtering by $p$ before a join gives the same rows as adding a column $k = p$, joining, and then keeping the rows where $k$ is true:
\[
[\,\mathsf{filter}(p);\ \mathsf{join}(T)\,] \ \equiv\ [\,\mathsf{withColumn}(k, p);\ \mathsf{join}(T);\ \mathsf{filter}(k);\ \mathsf{drop}(k)\,]
\]
The law extends to any downstream operators once they carry $k$ along. An aggregation, for example, ignores the rows whose $k$ is not true (Appendix~\ref{A-app:p3} lists the adaptation of every kind of downstream operator that \tool{} generates). Any well-typed predicate can fill $p$, so P3 needs no binding. The two sides stress the engine differently. On the first, fewer rows reach the downstream operators. On the second, all rows reach them, together with the column $k$ that nothing reads until the final filter. An optimizer rule such as column pruning, which removes columns that later operators do not use, must therefore keep $k$, and every downstream operator must carry it along correctly.

\textbf{P5: Predicate-Partition Decomposition.} Under SQL's three-valued logic, a predicate $p$ evaluates to exactly one of true, false, and null on each row. Filtering the output of a workload by $p$, by $\neg p$, and by $p$ \texttt{is null} therefore splits that output into three parts that together contain every row exactly once. For example, with $p := x > 0$, the rows with a positive $x$, the rows with a non-positive $x$, and the rows with a null $x$ make up the output. The split holds for any well-typed predicate, so P5 needs no binding. P5 tests how the engine evaluates predicates and how optimizer rules push them down or simplify them, which must keep every row in exactly one part.

\textbf{P7: Operator Input-Output Relations.} The output of an operator stands in a fixed relation to its input, whatever workload produced the input. For example, every row that $\mathsf{filter}(p)$ returns is a row of its input, so a binding can set $\mathrm{op} := \mathsf{filter}(p)$ and the relation to bag containment. Another binding states that intersecting a table with itself returns its distinct rows. These relations follow from the definitions of the operators. P7 tests each operator on inputs that come from generated multi-stage workloads, which exercise the plans that the optimizer builds around it.

\paragraph{Aggregate decomposition: P4.}
Many aggregates can be computed from partial results over parts of the data~\cite{gray1997datacube}. DISC engines rely on this: they compute partial aggregates per partition, shuffle them, and merge them.

\textbf{P4: Aggregation Decomposition.} An aggregate over all rows is related to the same aggregate computed per group. For example, the global count equals the sum of the per-group counts, so a binding sets $f := \mathsf{count}$ and $R(G, L) := G = \sum L$. Some aggregates, such as \texttt{avg}, cannot be computed from per-group results alone, so their relation is only a bound: the global average lies between the smallest and the largest per-group average (Figure~\ref{fig:motivating-instantiation}). Grouping by a generated key changes how the engine partitions and merges the work, so P4 tests that partial aggregation and merging agree with a direct computation.

\subsection{When Does an Oracle Hold?}
\label{sec:mp-laws}
A concrete property is a useful oracle only if it holds on a correct engine. This requires two things: the binding must be correct, that is, the premise on the left of $\Rightarrow$ in Table~\ref{tab:mp-overview} must hold (for P1, $e_1$ and $e_2$ must be equivalent), and the concrete property must then hold on every generated workload.

\paragraph{Correctness of bindings.} We do not formally prove that a binding is correct, because real engines have many semantic details that would be costly to model in full, such as the treatment of null and \texttt{NaN}, type coercion, ANSI mode~\cite{spark_ansi}, and session time zones. A developer therefore writes a binding from the documented semantics of the engine and first tests it in isolation on small inputs. Mistakes that remain show up as violations, which we triage.

\paragraph{From bindings to workloads.} If the binding is correct, the concrete property follows from the textbook principle that the meta-property builds on (Table~\ref{tab:mp-overview}). This argument assumes an idealized engine. In practice, most problems come from the following three sources.
\begin{itemize}[leftmargin=*,itemsep=2pt,topsep=3pt]
\item \emph{Nondeterminism.} The properties hold for deterministic workloads, that is, workloads that return the same bag of rows whenever they run on the same data. This assumption is common in metamorphic testing of database engines~\cite{rigger2020finding_tlp,norec_fse2020}. \tool{} meets it by construction: its generator (Section~\ref{sec:realization}) never emits nondeterministic functions such as \texttt{any\_value} or \texttt{sample}, and it uses order-dependent operators such as \texttt{limit} or \texttt{row\_number} only under a total order on exactly comparable columns.
\item \emph{Floating-point rounding.} Engines do not fix the order in which they process rows, and floating-point arithmetic~\cite{ieee754} depends on that order, so two workloads that compute the same sum in different orders can differ in its last digits. Checkers therefore compare floating-point values with a tolerance. This fix is partial, because a difference within the tolerance can grow beyond it when the value becomes a grouping or join key or is compared with another value. The generator therefore bounds the inputs of floating-point sums, averages, and variances and does not use their results as keys.
\item \emph{Overflow.} When a sum exceeds its result type, engines return a value that they define, such as a wrapped-around integer, infinity, or null for a decimal. The per-group sums of P4 may stay within range while the global sum overflows, so P4 can fail on a correct engine. \tool{} still generates such inputs, because they also trigger real bugs, and qualifies the assertion instead: the P4 checker also computes the sums exactly and reports no violation when one of them exceeds its result type. Under ANSI mode, the overflow raises an error, and \tool{} discards the run.
\end{itemize}
In general, \tool{} keeps generating boundary values, and when they expose such a case, we codify the outcome of triage once in the meta-property, either as a condition on its assertion or as a limit on what the generator produces, rather than once per test. In RQ1, triage was cheap: in one 12-hour campaign on each engine, 37 of the 194 violations were false alarms, fewer than one per 10{,}000 tests (\S\ref{sec:rq-new-bugs}).

\begin{figure}[t]
\centering
\small
\mprset{sep=1.2em}
\renewcommand{\arraystretch}{1.2}
\begin{tabular}{@{}l@{\quad}p{0.6\linewidth}@{}}
$\Sigma \vdash_s e : \tau$ & expression $e$ has type $\tau$ over schema $\Sigma$ and may appear at a site of class $s$ \\
$\vdash Q : \vS \Rightarrow \Sigma$ & workload $Q$ is schema-valid, from source schemas $\vS$ to sink schema $\Sigma$ \\
$\vdash C : (\Sigma_h \vdash_s \tau) \Rightarrow (\vS \Rightarrow \Sigma)$ & workload with one hole of class $s$, schema-valid once the hole holds an expression of type $\tau$ over $\Sigma_h$ \\
$\vdash \tst{W_1,\dots,W_m}{R} : \vS$ & test: schema-valid workloads over the same sources $\vS$ whose sink schemas $\Sigma_i$ fit the relation $R \subseteq \Bag{\Sigma_1} \times \dots \times \Bag{\Sigma_m}$ \\
\end{tabular}

\vspace{1ex}
\begin{mathpar}
\inferrule*[right=\rn{Seq}]{\vdash Q_1 : \vS \Rightarrow \Sigma_1 \\ \vdash Q_2 : \Sigma_1 \Rightarrow \Sigma_2}{\vdash [\,Q_1;\ Q_2\,] : \vS \Rightarrow \Sigma_2}
\and
\inferrule*[right=\rn{Plug}]{\vdash C : (\Sigma_h \vdash_s \tau) \Rightarrow (\vS \Rightarrow \Sigma) \\ \Sigma_h \vdash_s e : \tau}{\vdash C[e] : \vS \Rightarrow \Sigma}
\and
\inferrule*[right=\rn{Subst}]{\Gamma \vdash_s e : \tau \\ \Sigma_h \vdash \sigma : \Gamma}{\Sigma_h \vdash_s e\sigma : \tau}
\and
\inferrule*[right=\rn{Udf}]{f \text{ declared } \bar\tau \to \tau}{\bar x{:}\bar\tau \vdash_{\mathrm{row}} \mathsf{udf}_{\tau}(f)(\bar x) : \tau}
\end{mathpar}

\vspace{-0.5ex}
\noindent\rule{\linewidth}{0.3pt}
\vspace{-1ex}

\begin{mathpar}
\inferrule*[right=\rn{P1}]{
  \Gamma \vdash_s \bnd{e_1} : \tau \\ \Gamma \vdash_s \bnd{e_2} : \tau \\
  \vdash C : (\Sigma_h \vdash_s \tau) \Rightarrow (\vS \Rightarrow \Sigma) \\ \Sigma_h \vdash \sigma : \Gamma
}{\vdash \tst{C[e_1\sigma],\ C[e_2\sigma]}{\eqbag} : \vS}
\end{mathpar}

\vspace{-1ex}
\caption{Judgments and typing rules. The rule of P1 is shown as an example, and Appendix~\ref{A-app:rules} gives the rules of all seven meta-properties. Highlighted premises come from the binding, and the generator meets the rest. $\Sigma_h \vdash \sigma : \Gamma$ fills each generated hole in $\Gamma$ with a typed expression over $\Sigma_h$. $=$ compares outputs as bags, as in Table~\ref{tab:mp-overview}, with a tolerance for floating-point values (Section~\ref{sec:mp-laws}). Types ignore nullability.}
\label{fig:mp-rules}
\vspace{-2ex}
\end{figure}

\subsection{Type-Correct Bindings}
\label{sec:mp-typing}
\tool{} does not prove a binding correct (Section~\ref{sec:mp-laws}), but it checks mechanically that the binding, including the relation in its assertion, is type-correct. This check is weaker but automatic. It treats the holes as typed variables, much as a compiler checks a function body against the types of its parameters, so it needs no workload and runs once per binding, before any test. A binding that passes the check can be placed in any workload that \tool{} generates, and the resulting test is schema-valid: every column that it references exists with the expected type, every operator and expression gets inputs of the types that it needs, and the assertion compares outputs whose schemas match.\footnote{A schema-valid test can still fail at run time, for example with an overflow under ANSI mode, or break an engine rule that goes beyond types, which local validation catches during generation (Section~\ref{sec:realization-algorithm}).}

Figure~\ref{fig:mp-rules} states these checks as typing rules. A schema $\Sigma = c_1{:}\tau_1, \dots, c_n{:}\tau_n$ assigns a type to each column. The judgment $\vdash Q : \vS \Rightarrow \Sigma$ means that workload $Q$ is schema-valid and that its output has schema $\Sigma$. The judgment $\Sigma \vdash_s e : \tau$ means that expression $e$ has type $\tau$ and may appear at a site of class $s$. Sites are needed because engines also restrict where some expressions may appear. Spark, for example, accepts an aggregate function only in an aggregation and \texttt{explode} only in a projection. We omit the typing rules of individual operators and expressions, which follow the engine's documentation. A join, for example, concatenates the schemas of its inputs.

Checking the binding alone suffices because the typing rule of each meta-property divides its premises between the binding and the generator. A binding is type-correct if its holes have the types that the highlighted premises in Figure~\ref{fig:mp-rules} require. The other premises concern the surrounding workload, and the generator meets them by construction for every test (Section~\ref{sec:realization}). In P1, for example, the check verifies that $e_1$ and $e_2$ have the same type $\tau$, and the generator builds every workload around a hole of type $\tau$, the context $C$ in Figure~\ref{fig:mp-rules}, and fills the array $a$ and the predicate $p$ with columns of the right types. The following proposition makes this division precise.

\begin{proposition}[Schema validity]\label{prop:schema-valid}
Let the rule of a meta-property in Figure~\ref{fig:mp-rules} or Appendix~\ref{A-app:rules} conclude $\vdash \tst{W_1,\dots,W_m}{R} : \vS$. If the binding is type-correct and the generated workload satisfies the other premises, then each $W_i$ is schema-valid, that is, $\vdash W_i : \vS \Rightarrow \Sigma_i$ is derivable from the rules of Figure~\ref{fig:mp-rules} and the typing rules of the operators, and $R \subseteq \Bag{\Sigma_1} \times \dots \times \Bag{\Sigma_m}$.
\end{proposition}
\begin{proof}[Proof sketch]
By cases on the meta-property. In P1, \rn{Subst} gives $\Sigma_h \vdash_s e_i\sigma : \tau$ for $i = 1, 2$, and \rn{Plug} then gives $\vdash C[e_i\sigma] : \vS \Rightarrow \Sigma$. Both outputs have schema $\Sigma$, so $=$ compares bags of the same schema. P2 is analogous, with \rn{Udf} typing the UDF call. In P3 to P7, each workload extends a schema-valid prefix with operators whose typing rules the premises satisfy, so \rn{Seq} applies. The relation fits the output schemas by its premise in P4 and P7, and because all outputs have schema $\Sigma$ in P3, P5, and P6.
\end{proof}

We implement the check with the type inference of Spark's analyzer and Flink's planner, and all 203 bindings pass it.

\section{Concrete Property Realization}
\label{sec:realization}
In Section~\ref{sec:mp}, a meta-property together with a binding determines a concrete property. Parts 1 and 2 of Figure~\ref{fig:motivating-instantiation}, for example, give the property that \emph{the average of a column over all rows lies between the smallest and the largest per-group average}. A concrete property is still not an executable test. It needs a workload and data around it: the column that \texttt{avg} is applied to must exist in the output of the upstream workload and have a type that \texttt{avg} accepts. These surrounding parts should also vary from test to test rather than stay fixed, so that the same property is checked across many combinations of operators, schemas, and values. \tool{} \emph{realizes} a concrete property as executable tests: it generates the workload, schemas, and data anew for each test, and it builds them around the binding so that every test runs on the engine.

\subsection{Realization Algorithm}
\label{sec:realization-algorithm}
\renewcommand{\algorithmicrequire}{\textbf{Input:}}
\renewcommand{\algorithmicensure}{\textbf{Output:}}
\begin{algorithm}[h]
\caption{\textsc{Realize}: turning a concrete property into executable tests}
\label{alg:realize}
\small
\begin{algorithmic}[1]
\Require concrete property $(\mathsf{MP}, b)$: meta-property $\mathsf{MP}$ with binding $b$
\Ensure executable test: related workloads $W_1,\dots,W_m$ and an assertion $A$ on their outputs
\State $(G, r) \gets \Call{GenSkeleton}{\mathsf{MP}}$ \label{ln:skeleton} \Comment{fix dependencies and reserve binding site $r$}
\ForAll{source node $n \in G$} \label{ln:source-schema-for}
    \State $\Sigma[n] \gets \Call{GenSchema}{}$;\ \ $D[n] \gets \Call{GenData}{\Sigma[n]}$ \label{ln:source-schema}
\EndFor \label{ln:source-schema-end}
\ForAll{non-source $n \in \Call{TopoOrder}{G}$} \label{ln:topo} \Comment{dependency-ordered fill-in}
    \State $\Sigma_{\mathrm{in}} \gets \{\Sigma[m] : m \in \Call{Inputs}{n}\}$ \Comment{input schemas}
    \Repeat
        \If{$n = r$}
            \State $\mathit{Ops}[n] \gets \Call{PlaceBinding}{\mathsf{MP}, b, \Sigma_{\mathrm{in}}}$ \label{ln:place}
        \Else
            \State $\mathit{Ops}[n] \gets \Call{GenOperator}{\Sigma_{\mathrm{in}}}$ \label{ln:genop}
        \EndIf
    \Until{$\Call{IsWellFormed}{\mathit{Ops}[n], \Sigma_{\mathrm{in}}}$ \textbf{or} retries exhausted} \label{ln:until}
    \State $\Sigma[n] \gets \Call{InferSchema}{\mathit{Ops}[n], \Sigma_{\mathrm{in}}}$ \label{ln:update-schema} \Comment{schema inference}
\EndFor
\State $\langle W_1,\dots,W_m\rangle \gets \Call{EmitWorkloads}{G, \mathit{Ops}}$ \label{ln:emit}
\State $A \gets \Call{EmitAssertion}{\mathsf{MP}, b}$ \label{ln:law} \Comment{asserts $R(\sem{W_1},\dots,\sem{W_m})$}
\State \Return $(\langle W_1,\dots,W_m\rangle, A)$
\end{algorithmic}
\vspace{1pt}
{\scriptsize\raggedright
$G$ is a workload skeleton, and $r$, the \emph{binding site}, is the node reserved for the binding. \textsc{Inputs}$(n)$ is the set of nodes that $n$ reads from. $\Sigma$
holds the schema of each node and $D$ the data of each source. $\mathit{Ops}[n]$
is the operator generated for node $n$ or, at $r$, the related operators that
the meta-property needs, such as the two sides of a rewrite. $R$ is the property of
$\mathsf{MP}$ in Table~\ref{tab:mp-overview}, and $\sem{W_i}$ is the output of
$W_i$ on the generated data. \textsc{GenOperator},
\textsc{PlaceBinding}, and \textsc{InferSchema} follow the typing rules of
Section~\ref{sec:mp-typing}. \textsc{IsWellFormed} applies local validation, which also catches operators that the engine's analyzer rejects. If a node exhausts its retries,
the attempt is abandoned and generation restarts. \par}
\end{algorithm}

Algorithm~\ref{alg:realize} describes how \tool{} turns a concrete property into an executable test. It makes random choices, so running it many times on the same concrete property yields many different tests. It proceeds in four phases. (1)~It samples a workload skeleton (line~\ref{ln:skeleton}), which fixes the DataFrame nodes of the workload and, for each node, the nodes that it reads from. It also reserves the \emph{binding site} $r$, the node where the binding goes, for example the node where an equivalence rewrite takes place. (2)~It gives each source node a schema and data (lines~\ref{ln:source-schema-for}--\ref{ln:source-schema-end}). (3)~It fills the other nodes in topological order (lines~\ref{ln:topo}--\ref{ln:update-schema}). The binding site receives the binding (line~\ref{ln:place}), and every other node receives a generated operator (line~\ref{ln:genop}). \tool{} checks each filled node before it moves on and infers the node's output schema (lines~\ref{ln:until}--\ref{ln:update-schema}). (4)~It emits the related workloads of the test and an assertion that their outputs satisfy the property (lines~\ref{ln:emit}--\ref{ln:law}). \tool{} then runs the workloads on the engine and reports the test if the assertion fails. Meta-properties differ only in which nodes may serve as the binding site (line~\ref{ln:skeleton}), for example only the end of the workload for P4, in how the binding is placed (line~\ref{ln:place}), and in which property the assertion checks (line~\ref{ln:law}). We next explain its core design choices.

\paragraph{Schema-aware.} In a DataFrame workload, every operator changes the schema that later operators see. A generator that ignores schemas soon picks an expression whose operands have incompatible types, refers to a column that an earlier projection removed, or joins two inputs that do not fit, and the engine rejects the plan before it tests anything. \tool{} therefore keeps the schema of every node, which is exactly the typing judgment $\vdash Q : \vS \Rightarrow \Sigma$ of the previous section. At each node, \textsc{GenOperator} picks an operator and its arguments, such as the columns and constants of a join predicate, that are compatible with the schemas of the node's inputs (line~\ref{ln:genop}). \textsc{InferSchema} then infers the node's output schema (line~\ref{ln:update-schema}),\footnote{Schema inference is static: \tool{} infers each schema while it generates the workload, without running the engine. In 1{,}700 generated programs, the schemas that \textsc{InferSchema} infers for 9{,}167 DataFrames match those that Spark reports in every column name and base type.} which the nodes that read from it use in turn.

\paragraph{Dependency-ordered.} A node can be filled only after the schemas of its inputs are known. Visiting the nodes in topological order guarantees this, in the same way that the rule \rn{Seq} composes typed workload fragments. This order is what makes schema-aware generation work for multi-stage workloads and not only for single operators.

\paragraph{Binding-guided.} One could generate a whole workload first and then search it for a place where the binding fits, but such a place rarely exists, because the columns that the binding needs must be visible there and the surrounding operator must accept the binding's result type. \tool{} instead reserves the binding site before it fills the skeleton. When it reaches that node (line~\ref{ln:place}), it fills the generated holes with visible columns of the required types and chooses a host operator that accepts the binding's result, which are the premises of the rules \rn{Plug} and \rn{Subst}.

\paragraph{Local validation.}\label{sec:filling-operators-bindings}\label{sec:realization-checkers} Schema-aware generation does not guarantee that the engine accepts a workload, because engines add requirements of their own beyond types. In Spark, for example, joining $df_a$ and $df_b$ with the trivial predicate $df_a.x = df_a.x$ looks harmless, since the predicate refers only to $df_a$. The analyzer, however, may resolve it as $df_a.x = df_b.x$, and the workload fails if $df_b$ has no column $x$. \tool{} checks each node for such known cases as soon as it fills the node (line~\ref{ln:until}) and resamples the node on failure, retrying five times by default before it restarts the workload. Rejecting a node during generation is cheap, whereas a rejected workload is found only after the engine has started to run it. The checks are not complete, but this is harmless: a workload that slips through fails analysis at run time, and \tool{} discards it without adding to the triage effort.

\begin{figure}[t]
\centering
\footnotesize
\resizebox{\linewidth}{!}{%
\begin{tikzpicture}[
    x=1cm, y=1cm,
    df/.style={draw, rounded corners=2pt, fill=gray!6, align=left, inner sep=3pt},
    res/.style={draw, rounded corners=2pt, fill=bindbg, align=left, inner sep=3pt},
    outbox/.style={draw, dashed, rounded corners=2pt, align=left, inner sep=4pt},
    sch/.style={font=\scriptsize\color{gray!70!black}, align=left},
    ph/.style={circle, draw, inner sep=0.6pt, font=\scriptsize\bfseries},
    arr/.style={->, >=stealth, thick, gray!70!black},
    emit/.style={-{Implies}, double, double distance=1.2pt, thick, gray!80!black}
]
\node[df] (a) at (0, 0.7) {\texttt{a}: \texttt{x:int}\\[-1pt]{\scriptsize [\texttt{INT\_MAX}, \texttt{INT\_MAX}, 1]}};
\node[df] (b) at (0,-0.7) {\texttt{b}: \texttt{y:int}\\[-1pt]{\scriptsize $[1,\ 2]$}};
\node[df] (n) at (3.5, 0) {\texttt{n = a.crossJoin(b)}};
\node[sch, below=1pt of n] {inferred: \texttt{x:int, y:int}};
\node[res] (r) at (8.1, 0) {binding site $r$\\[1pt]
  $W_1$: \texttt{agg(avg(x))}\\[-1pt]
  $W_2$: \texttt{groupBy(x).agg(avg(x))}};
\node[sch, below=1pt of r] {\texttt{x}: an \texttt{int} column of \texttt{n},\\[-1pt]used as value and key};
\node[outbox] (o) at (13.2, 0) {$G$ = output of $W_1$\ \ {\scriptsize\color{gray!70!black}(\texttt{g:int})}\\[1pt]
  $L$ = output of $W_2$\ \ {\scriptsize\color{gray!70!black}(\texttt{x:int, l:int})}\\[3pt]
  \colorbox{bindbg}{assert $\min L \le G \le \max L$}};

\draw[arr] (a) -- (n);
\draw[arr] (b) -- (n);
\draw[arr] (n) -- (r);

\node[ph] at (-1.6, 1.6) {1};
\node[anchor=west, font=\scriptsize] at (-1.4, 1.6) {skeleton: sources \texttt{a}, \texttt{b} $\to$ node \texttt{n} $\to$ binding site $r$};
\node[ph] at (-1.6, -1.55) {2};
\node[anchor=west, font=\scriptsize] at (-1.4, -1.55) {schemas and data of the sources};
\node[ph] at (1.9, -1.15) {3};
\node[anchor=west, font=\scriptsize] at (2.1, -1.15) {fill \texttt{n}, then place the binding at $r$};
\node[ph] at (11.2, 1.1) {4};
\node[anchor=west, font=\scriptsize] at (11.4, 1.1) {emit the workloads, run, and check};
\end{tikzpicture}%
}
\caption{Realizing the concrete property of Figure~\ref{fig:motivating-instantiation} as an executable test. \tool{} generates the gray parts and infers their schemas. The highlighted parts come from the binding: the two workloads $W_1$ and $W_2$ differ only at the binding site $r$, and the assertion checks the property on their outputs. Numbers mark the phases of Algorithm~\ref{alg:realize}.}
\label{fig:realize-example}
\end{figure}

\medskip\noindent\textbf{Example.} Figure~\ref{fig:realize-example} shows how Algorithm~\ref{alg:realize} realizes the test of the motivating example in Figure~\ref{fig:motivating-instantiation}. (1)~First, the skeleton is fixed: two sources, \texttt{a} and \texttt{b}, a node \texttt{n} that reads both, and the binding site $r$ at the end of the workload, where P4 places its aggregations. (2)~Next, the sources receive schemas and data. For example, \texttt{a} has an integer column \texttt{x} with the values \texttt{INT\_MAX}, \texttt{INT\_MAX}, and 1, where \texttt{INT\_MAX} is the largest \texttt{INT}. (3)~Then the nodes are filled in dependency order. For \texttt{n}, \textsc{GenOperator} picks a cross join, and \textsc{InferSchema} infers the output schema \texttt{x:int, y:int}. At the binding site $r$, \textsc{PlaceBinding} needs a value column whose type \texttt{avg} accepts and a grouping key, and it picks \texttt{x} for both. It then places the two operators that P4 requires, \texttt{agg(avg(x))} for the global average and \texttt{groupBy(x).agg(avg(x))} for the per-group averages. (4)~Finally, \tool{} emits two workloads that are identical except at the binding site, where one has the global aggregation and the other the grouped one, together with an assertion that the global average $G$ lies between the smallest and the largest per-group average in $L$.

\subsection{Skeletons, Schemas, and Data}
\label{sec:workload-skeletons}
\label{sec:source-schema-data}
This subsection details the skeleton samplers and the data generator that phases (1) and (2) use. A skeleton is a DAG whose nodes are source or intermediate DataFrames and whose edges are data dependencies. A non-source node is later filled with a unary operator, such as \texttt{filter}, \texttt{select}, or \texttt{withColumn}, or a binary operator, such as \texttt{join}, \texttt{union}, \texttt{intersect}, or \texttt{except}. \tool{} has two skeleton samplers.
\begin{itemize}
    \item \textsc{TreeGen}$(depth, r_{\mathrm{bin}})$ samples a tree backward from the sink. Its leaves are sources, $depth$ bounds the depth with leaves at depth 1, and $r_{\mathrm{bin}}$ is the probability that an internal node is binary.
    \item \textsc{DAGGen}$(n_0,n_1,n_2)$ samples a DAG with $n_0$ sources, $n_1$ unary nodes, and $n_2$ binary nodes.
\end{itemize}
By design, both samplers produce skeletons with a single sink, the node whose output the test compares. Every node then lies on a path to that sink, so no generated operator is wasted on an output that no check observes.

Each source gets an independent schema of 3--5 columns by default. Column types are drawn from eight top-level types, \texttt{int}, \texttt{float}, \texttt{string}, \texttt{bool}, \texttt{date}, \texttt{timestamp}, \texttt{array}, and \texttt{struct}, whose elements and fields are primitive. Rows are then generated value by value from the column types, and each source has 0--10 rows by default. A \texttt{float} value, for example, comes from a numeric range, but with 5\% probability it is a special value such as \texttt{NaN}, \texttt{Inf}, or \texttt{null}. Other types have their own samplers, which also pick boundary values with some probability, for example the smallest and largest \texttt{INT}, $-0.0$ and the largest \texttt{FLOAT}, the 2038 overflow of 32-bit timestamps, leap days, and times that a daylight-saving change skips or repeats. Such values often make arithmetic overflow. Under ANSI mode, the engine then raises an error and \tool{} discards the run. Otherwise, the engine returns a value that it defines, which the checks handle as described in the previous section. \tool{} still generates these values on purpose, because they also trigger real bugs.

\subsection{Implementation}

\begin{table}[t]
\caption{Implementation size in lines of code (LoC). One workload and data
generator is shared by Spark and Flink; each engine adds only an adapter, a
template per meta-property, and bindings. Average binding LoC is per concrete
property and excludes P3 and P5, which need no binding. P1 is Spark-only.}
\label{tab:generality}
\centering
\small
\setlength{\tabcolsep}{2.5pt}
\begin{tabular}{l rrrr rrrr}
\toprule
 & \multicolumn{4}{c}{\textbf{Spark}} & \multicolumn{4}{c}{\textbf{Flink}} \\
\cmidrule(lr){2-5}\cmidrule(lr){6-9}
\multicolumn{9}{@{}l}{\textit{Base}} \\
\quad Workload and data generator & \multicolumn{8}{c}{9{,}922 (shared by both engines)} \\
\quad Engine adapter and runner & \multicolumn{4}{c}{2{,}723} & \multicolumn{4}{c}{930} \\
\midrule
 & \textbf{MP} & & \multicolumn{2}{c}{\textbf{Binding LoC}}
 & \textbf{MP} & & \multicolumn{2}{c}{\textbf{Binding LoC}} \\
\cmidrule(lr){4-5}\cmidrule(lr){8-9}
 & \textbf{LoC} & \textbf{\#Props} & \textbf{Total} & \textbf{Avg.}
 & \textbf{LoC} & \textbf{\#Props} & \textbf{Total} & \textbf{Avg.} \\
\midrule
P1 & 583 & 56 & 923     & 16.5 & --  & -- & --  & --   \\
P2 & 818 & 54 & 1{,}164 & 21.6 & 314 & 15 & 393 & 26.2 \\
P3 & 709 & 1  & --      & --   & 44  & 1  & --  & --   \\
P4 & 212 & 24 & 358     & 14.9 & 155 & 5  & 93  & 18.6 \\
P5 & 65  & 1  & --      & --   & 79  & 1  & --  & --   \\
P6 & 182 & 10 & 105     & 10.5 & 77  & 2  & 20  & 10.0 \\
P7 & 197 & 24 & 788     & 32.8 & 69  & 13 & 176 & 13.5 \\
\midrule
Total & 2{,}766 & 170 & 3{,}338 & 19.9 & 738 & 37 & 682 & 19.5 \\
\bottomrule
\end{tabular}
\end{table}

\label{sec:portability}
The engineering effort of \tool{} lies in two parts, the base infrastructure of the generators and the parts specific to the meta-properties (Table~\ref{tab:generality}). The workload and data generator has 9{,}922 lines and is shared by both engines. Each engine adds an adapter and a runner, 2{,}723 lines for Spark and 930 for Flink. The Spark adapter generates programs with the DataFrame API, and the Flink adapter with the PyFlink Table API in batch mode. Because DISC engines expose similar typed relational APIs, we expect that supporting another engine mostly requires mapping these API names. The parts specific to the meta-properties are decoupled from the generator. A meta-property needs a template of 44 to 818 lines per engine, and a concrete property needs only a binding, of 19.9 lines on average on Spark and 19.5 on Flink. In total, our prototype has 168 bindings on Spark and 35 on Flink, which give 207 concrete properties, 170 on Spark and 37 on Flink, when P3 and P5, which need no binding, are counted once per engine. Flink lacks only P1 (Higher-Order Expression Equivalence Rewrite), because it has no higher-order functions with lambda arguments.

\section{Evaluation}
We evaluate \tool{} through four research questions:
\begin{enumerate}
\item[\textbf{RQ1}] \textbf{Real-World Bug Detection:} Can \tool{} find previously unknown correctness bugs in the latest releases of real DISC systems?
\item[\textbf{RQ2}] \textbf{Generality on Real-World Bugs:} How expressive are \tool{}'s meta-properties, and how many known real-world correctness bugs do they cover?
\item[\textbf{RQ3}] \textbf{Coverage of Generated Workloads:} Do \tool{}'s generated workloads exercise more of the engine than those of existing testing tools?
\item[\textbf{RQ4}] \textbf{Validity and Scalability:} Does generation remain valid as schema width, data volume, and workload complexity grow?
\end{enumerate}

\begin{table}[tp]
\centering
\caption{Seventeen previously unknown wrong-result bugs that \tool{} found in
the latest releases of two mature frameworks, Spark 4.2.0 and Flink 2.3.0.
Each bug was exposed by a concrete property, instantiated from a short binding
or, for P3 and P5, from the meta-property alone. We list each bug with its
meta-property and root cause. Every meta-property exposed at least one bug.
None of the bugs crashes, so tools that detect only crashes cannot find them.}
\label{tab:new-bugs}
\scriptsize
\setlength{\tabcolsep}{4pt}
\renewcommand{\arraystretch}{1.05}
\newcommand{\rc}{\textbf{Root cause:} }
\begin{tabularx}{\textwidth}{@{}l >{\raggedright\arraybackslash}X@{}}
\toprule
\textbf{Issue} & \textbf{Property that exposed the bug, and root cause} \\
\midrule
Spark-1 & P7 (Operator Input-Output Relations): filtering by \texttt{EXISTS}($S$) returns the rows of a semi join with $S$. \\
         & \rc After a rename, an outer reference binds to the inner plan, so the correlation always holds. \\[2pt]
Spark-2 & P2 (UDF Counterpart Equivalence Rewrite): an identity Python UDF returns a timestamp unchanged. \\
         & \rc PySpark loses the daylight-saving \texttt{fold} flag, so a repeated-hour time shifts by an hour. \\[2pt]
Spark-3 & P1 (Higher-Order Expression Equivalence Rewrite): \texttt{array\_max(arr)} equals the maximum over \texttt{explode(arr)}. \\
         & \rc A rule drops a \texttt{distinct} that normalized $-0.0$ to $0.0$, so keys from one array mismatch. \\[2pt]
Spark-4 & P1 (Higher-Order Expression Equivalence Rewrite): equivalent expressions agree, compiled or interpreted. \\
         & \rc Interpreted \texttt{max}/\texttt{min} over collated strings keep a reference to a reused buffer, not a copy. \\[2pt]
Spark-5 & P5 (Predicate-Partition Decomposition): the rows where $p$ is true, false, or \texttt{NULL} make up the input. \\
         & \rc A filter pushed below a \texttt{distinct} sees the raw $-0.0$, while the \texttt{distinct} outputs $0.0$. \\[2pt]
Spark-6 & P6 (Intermediate Materialization Equivalence): an intermediate ORC round trip keeps the output. \\
         & \rc ORC filter pushdown compares timestamps in two calendars and skips rows before 1582. \\[2pt]
Spark-7 & P6 (Intermediate Materialization Equivalence): checkpointing an intermediate result keeps the output. \\
         & \rc Decorrelation computes an outer \texttt{DOUBLE} sum in two ways, so the join back misses. \\[2pt]
\midrule
Flink-1 & P4 (Aggregation Decomposition): the global average lies within the range of the per-group averages. \\
         & \rc \texttt{AVG} is rewritten into \texttt{SUM}/\texttt{COUNT} below a join, and the \texttt{SUM} keeps the \texttt{INT} type and overflows. \\[2pt]
Flink-2 & P7 (Operator Input-Output Relations): intersecting a table with itself returns its distinct rows. \\
         & \rc The planner treats a \texttt{ROLLUP} output as unique on its keys and drops the deduplication. \\[2pt]
Flink-3 & P6 (Intermediate Materialization Equivalence): an intermediate file-system round trip keeps the output. \\
         & \rc $x \geq x$ is simplified to $x$ \texttt{IS NOT NULL}, which is wrong when $x$ is NaN. \\[2pt]
Flink-4 & P3 (Predicate Materialization with Downstream Adaptation): filtering by $p$ early equals tagging rows with $p$ and filtering at the end. \\
         & \rc A \texttt{DISTINCT} pushed below a cross join makes an empty grouped aggregate return one row. \\[2pt]
Flink-5 & P4 (Aggregation Decomposition): the global sum equals the sum of the per-group sums. \\
         & \rc An overflowing \texttt{DECIMAL} sum is treated as ``no value yet'' and restarts from the next row. \\[2pt]
Flink-6 & P4 (Aggregation Decomposition): the global minimum equals the minimum of the per-group minima. \\
         & \rc \texttt{MIN}/\texttt{MAX} never adopts or replaces NaN, so the result depends on whether NaN comes first. \\[2pt]
Flink-7 & P5 (Predicate-Partition Decomposition): the rows where $p$ is true, false, or \texttt{NULL} make up the input. \\
         & \rc \texttt{BETWEEN} with equal bounds becomes an equality that drops the bound's trailing spaces. \\[2pt]
Flink-8 & P4 (Aggregation Decomposition): the global maximum equals the maximum of the per-group maxima. \\
         & \rc \texttt{VAR\_SAMP}/\texttt{STDDEV\_SAMP} are typed non-nullable, so a single-row group's \texttt{NULL} becomes $-1$. \\[2pt]
Flink-9 & P7 (Operator Input-Output Relations): the union of a table with itself returns its distinct rows. \\
         & \rc $-0.0$ and $0.0$ are treated as different keys in grouping, deduplication, and hash joins. \\[2pt]
Flink-10 & P4 (Aggregation Decomposition): the global sum equals the sum of the per-group sums. \\
         & \rc \texttt{SUM} and \texttt{AVG} on \texttt{DECIMAL(38,$s$)} over a join are rounded to scale 6. \\[2pt]
\bottomrule
\end{tabularx}
\end{table}

\subsection{RQ1: Real-World Bug Detection}
\label{sec:rq-new-bugs}

\paragraph{Setup.}
We ran \tool{} on the latest releases of two DISC engines, Apache Spark 4.2.0 (PySpark, local mode) and Apache Flink 2.3.0 (PyFlink, batch mode). Spark campaigns use all seven meta-properties. Flink campaigns use all but P1 (Higher-Order Expression Equivalence Rewrite), because Flink SQL has no lambda functions. The Spark campaigns ran for 36 hours in total (about 1.7 million tests) and the Flink campaign for 12 hours (63{,}388 tests), each with six parallel workers on one server (AMD EPYC 9655P, 377\,GB of memory). For every property violation, we minimized the workload into a standalone reproducer, checked that it reproduces on the latest release, and searched the issue tracker for duplicates before reporting it.

\paragraph{Results.}
\tool{} exposed 17 previously unknown wrong-result bugs, 7 in Spark and 10 in Flink (Table~\ref{tab:new-bugs}).\footnote{All bugs were reported to the Apache JIRA. To comply with the double-blind review policy, we replace the issue identifiers with Spark-$i$ and Flink-$i$.} As we show at the end of this section, existing SQL and DISC testing tools could not have found them. Every meta-property exposes at least one bug, and P4 (Aggregation Decomposition) exposes the most (five). Most bugs need a specific upstream workload, such as a join below an aggregate (Flink-1, Flink-10), a \texttt{ROLLUP} over a nullable key (Flink-2), or a column rename before a correlated subquery (Spark-1). On a plain source table without such a workload, the same properties pass. \tool{} finds these bugs because its generator composes well-formed workloads around each binding and thus checks the same property in many contexts.

Section~\ref{sec:motivating} walks through Flink-1. We describe three more bugs below, each with its minimized reproducer, in which the code that the property adds is highlighted. We write a reproducer in SQL where it is shorter than the generated PySpark or PyFlink program.

\textbf{Case study: Flink-2.} This bug was exposed by P7 (Operator Input-Output Relations): intersecting a table with itself must return its distinct rows. In the workload below, a \texttt{ROLLUP} over a key \texttt{k} that is always \texttt{NULL} produces two identical rows $(\texttt{NULL}, 2)$, one for the \texttt{NULL} group and one for the grand total. Their self-intersection should return one row, but Flink returns both:
\begin{lstlisting}[style=pyspark,language=SQL,numbers=none,frame=none,xleftmargin=0pt,basicstyle=\ttfamily\footnotesize,escapeinside={(*@}{@*)}]
CREATE VIEW di AS SELECT k, COUNT(x) AS c
  FROM (VALUES (CAST(NULL AS INT), 1), (CAST(NULL AS INT), 2)) AS v(k, x)
  GROUP BY ROLLUP(k);                                  -- (NULL, 2), (NULL, 2)
SELECT * FROM di (*@\bndt{INTERSECT SELECT * FROM di}@*);       -- (NULL, 2), (NULL, 2)
\end{lstlisting}
Flink assumes that the \texttt{ROLLUP} output is unique on \texttt{k}, although it is unique only together with the grouping-set identifier, which is projected away. It therefore turns \texttt{INTERSECT} into a semi join without deduplication.

\textbf{Case study: Flink-3.} This bug was exposed by P6 (Intermediate Materialization Equivalence): writing an intermediate result to a table and reading it back must not change the final output. In the workload below, view \texttt{pre} copies a \texttt{DOUBLE} column that holds NaN into \texttt{a} and \texttt{b}, and the query keeps the rows where \texttt{a >= b}. The NaN row survives before materialization but not after:
\begin{lstlisting}[style=pyspark,language=SQL,numbers=none,frame=none,xleftmargin=0pt,basicstyle=\ttfamily\footnotesize,escapeinside={(*@}{@*)}]
CREATE VIEW pre AS SELECT x AS a, x AS b FROM (VALUES (CAST('NaN' AS DOUBLE)), (1.5)) s(x);
SELECT * FROM pre WHERE a >= b;      -- (NaN, NaN), (1.5, 1.5)
(*@\bndt{INSERT INTO m SELECT * FROM pre;}@*)   -- m: a file-system table
SELECT * FROM (*@\bndt{m}@*) WHERE a >= b;        -- (1.5, 1.5)
\end{lstlisting}
Before materialization, Flink sees that both operands are the same column and simplifies \texttt{a >= b} to \texttt{a IS NOT NULL}, which keeps NaN. After materialization, \texttt{a} and \texttt{b} are separate columns and are compared with IEEE semantics, under which NaN $\geq$ NaN is false.

\textbf{Case study: Spark-2.} This bug was exposed by P2 (UDF Counterpart Equivalence Rewrite): an identity Python UDF must return its input unchanged. In the workload below, the session time zone is America/Los\_Angeles, and \texttt{t} lies in the hour that repeats when daylight saving time ends. The UDF returns a time one hour earlier:
\begin{lstlisting}[style=pyspark,numbers=none,frame=none,xleftmargin=0pt,basicstyle=\ttfamily\footnotesize,escapeinside={(*@}{@*)}]
# session time zone America/Los_Angeles, t = 2021-11-07 01:30 PST (the repeated hour)
df.select((*@\bndt{F.udf(lambda x: x, 'timestamp')('t')}@*))   # 2021-11-07 01:30 PDT, one hour earlier
df.select((*@\bndt{F.col('t')}@*))                               # 2021-11-07 01:30 PST
\end{lstlisting}
Python marks the second occurrence of a repeated local time with a \texttt{fold} flag, and PySpark drops this flag when it passes timestamps between the JVM and Python. Since Spark 4.2 makes Arrow-optimized Python UDFs the default, a UDF that reads such a timestamp can receive the wrong instant.

\paragraph{False alarms.}
Some properties hold only in exact arithmetic or without overflow, so a violation is not always a bug. As Section~\ref{sec:mp-laws} explains, \tool{} still generates boundary values such as NaN, $-0.0$, and extreme numbers, because they also trigger real bugs, and we triage the violations. In a 12-hour Spark campaign (498{,}333 tests) and the Flink campaign, 157 of the 194 violations are real bugs, new or known. The other 37 are false alarms, 16 in Spark and 21 in Flink, fewer than one per 10{,}000 tests: fifteen come from floating-point rounding, eighteen from overflow, and four from defects in \tool{}'s type inference for \texttt{COUNT} and \texttt{AVG}, which we fixed. Each class needed only one fix, so triage effort does not grow with the number of tests.\footnote{We fixed the type inference. For rounding, the generator bounds the inputs of floating-point sums, averages, and variances, and does not use their results as keys. Except for one UDF binding that returned values outside the \texttt{INT} range and now returns \texttt{NULL} instead, overflow occurred where P4 compares a \texttt{SUM} or \texttt{AVG} over the whole input with the same aggregate per group. Its oracle now also computes these sums exactly in Python and does not report a violation if any of them exceeds the range of the result type.}

\paragraph{Could existing tools find these bugs?}
Tools that detect only crashes or errors cannot detect wrong results. CometFuzz~\cite{cometfuzz_github} tests Spark differentially, with and without the Comet accelerator~\cite{datafusion_comet}, but each of the seven Spark bugs lies either in Spark's analyzer or optimizer, which produces the plan for both runs, or in a path where Comet falls back to Spark (Python UDFs, collated strings, and ORC scans), so both runs share it. SQLancer proposes oracles that check relations between queries, such as TLP~\cite{rigger2020finding_tlp}, NoREC~\cite{norec_fse2020}, and PQS~\cite{pqs_osdi2020}. TLP resembles P5 (Predicate-Partition Decomposition), and its aggregate variant resembles P4 (Aggregation Decomposition) with three chunks defined by a predicate. NoREC resembles P3 (Predicate Materialization with Downstream Adaptation): both move a predicate from a filter into a computed column. Even in principle, nine of the 17 bugs need relations that none of these oracles states.\footnote{TLP could express Spark-5, Flink-1, Flink-5, Flink-6, Flink-7, Flink-9, and Flink-10, and NoREC could express Flink-3. The other nine need a UDF and its builtin counterpart (Spark-2), equivalent higher-order expressions or evaluation modes (Spark-3, Spark-4), predicate materialization (Flink-4), intermediate materialization (Spark-6, Spark-7), a sample variance (Flink-8), or operator input-output relations (Spark-1, Flink-2).} In practice, SQLancer could find none of the 17 bugs. It has no Flink provider, and its Spark provider has only limited TLP support~\cite{sqlancer_spark_pr}.\footnote{Spark-5 is the only Spark bug that TLP could express, but it needs a \texttt{DISTINCT}, which the provider cannot generate.}

\begin{table}[t]
\centering
\caption{
\textbf{Expressiveness of meta-properties} on all 178 fixed Spark
correctness issues filed from 2020 to 2025. In-scope issues are deterministic
wrong results in core query optimization and computation; out-of-scope issues
lie outside it, in data sources, other front ends, streaming or
nondeterminism, or produce no wrong query result. Of the 65 testable in-scope
issues, the meta-properties express 55, and from the binding alone, \tool{}
rediscovers 26 of them within 600 generated workloads. \tool{} also found the 17 previously unknown bugs in Table~\ref{tab:new-bugs}.
} 
\label{tab:bug-census}
\small
\newcommand{\lone}{\hspace*{1.0em}}
\newcommand{\ltwo}{\hspace*{2.0em}}
\begin{tabular}{@{}l r@{}}
\toprule
\textbf{Category} & \textbf{\#Issues} \\
\midrule
\textbf{In scope: core query optimization and computation} & \textbf{71} \\
\lone \textbf{Covered by P1--P7} & \textbf{55} \\
\ltwo Rediscovered end to end (within 600 workloads) & 26 \\
\lone Not covered by P1--P7 & 10 \\
\lone Not testable on a release & 6 \\
\midrule
\textbf{Out of scope: outside core computation} & \textbf{107} \\
\lone Data sources, file formats, and catalog & 49 \\
\lone Other front ends (SQL-only syntax, typed Scala/Java APIs, Spark Connect) & 26 \\
\lone Streaming, cluster-only, or nondeterministic & 18 \\
\lone No wrong query result & 14 \\
\midrule
\textbf{Total} & \textbf{178} \\
\bottomrule
\end{tabular}
\end{table}
\subsection{RQ2: Generality on Real-World Bugs}
\label{sec:rq-generality}

\paragraph{Setup.}
We study all 178 Spark issues that are labeled \emph{correctness}, which marks silently wrong results, resolved as fixed, and created between 2020 and 2025~\cite{spark_jira_correctness}. An issue is in scope if it is a deterministic wrong-result bug that a PySpark DataFrame workload over generated in-memory data can trigger. For each in-scope issue, we ask two questions. First, is it \emph{covered} by our meta-properties: does some binding yield a concrete property that fails on a buggy release and holds on the fixed release, given the issue's triggering input? Second, is it \emph{found end to end}: given only that binding, without the issue's data or workload, does a \tool{} campaign expose the bug within 600 generated workloads, with a violating workload that passes on the fixed release? Claude Opus 5.5 classified the issues and wrote the bindings, and an author reviewed the classification and the coverage decisions. The end-to-end campaigns run automatically with the generator of RQ1, without overfitting it to the known bugs.

\paragraph{Which bugs are out of scope?}
107 of the 178 issues are outside \tool{}'s scope (Table~\ref{tab:bug-census}). (1)~35 involve data sources, file formats, or JDBC, and 14 involve catalog, DDL, or table caching, while \tool{} generates workloads over in-memory data. (2)~26 are reachable only through other front ends: SQL-only syntax, the typed Scala and Java APIs, or the Spark Connect clients. These layers translate user code into the same logical plans as PySpark, and \tool{} exercises the optimizer and execution engine behind them, not the layers themselves. (3)~10 concern Structured Streaming, and 8 need a multi-executor cluster, concurrent jobs, or a nondeterministic query, which \tool{}'s deterministic local campaigns exclude by design. (4)~14 do not show a wrong result of a DataFrame query on Spark itself, for example an MLlib model, a UI metric, an error that Spark fails to raise, a JVM crash, or a result that is wrong only with a third-party backend. Of the 71 in-scope issues, we could not test six, which were introduced and fixed between two releases and so never appeared in a release.

\paragraph{Which bugs can the meta-properties not express?}
The meta-properties cover 55 of the 65 testable issues. Of the other ten, four concern values that Spark treats as equal, such as $-0.0$ and $0.0$ or two NaNs, and three depend on the physical execution strategy, such as the join algorithm that a hint selects. They suggest two new meta-properties: a result stays the same when a value is replaced by an equal one, and when the engine executes the same workload with a different physical strategy. Adding a meta-property is relatively cheap, because the generator stays unchanged. It needs only a template, which realizes the meta-property in generated workloads, and a few concrete bindings. The existing templates have a median of 182 lines, and a binding has under 20 lines on average (Table~\ref{tab:generality}).

\paragraph{Why are some covered bugs not found end to end?}
Testing does not guarantee that a bug is exposed. \tool{} does not fix the upstream workload or the data, so it must generate the inputs that trigger a bug. Of the 29 covered bugs that it misses within 600 workloads, 16 need a specific combination of operators, such as a \texttt{USING} join or two window functions over the same window, 8 need specific values, such as a whole-\texttt{NULL} struct or a regular expression that matches the empty string, and 2 need large inputs. The same freedom, however, lets \tool{} reach a bug through workloads that its reporter never wrote. From the binding alone, it finds the other 26 covered bugs after a median of 6.5 workloads (10.5 seconds), and in 25 of them the detecting workload differs from the issue's reproducer in its values, its shape, or both. For example, the reporter of SPARK-31500 inserted the same bytes twice to make \texttt{collect\_set} return duplicates, whereas \tool{} produced the duplicates with an \texttt{explode}. A binding thus works as a regression test that extends the single trigger of a bug report to many workloads.

\begin{table}[t]
\centering
\small
\caption{
Validity and diversity of generated queries per tool on Spark
4.2.0. A query is plan-valid if it passes Spark's logical-plan stage without
error and runtime-valid if it also executes without error. \tool{}'s queries
are 28--37 points more often valid than CometFuzz's and cover far more
distinct physical operators, expressions, and operator trees.
}
\vspace{-3ex}
\label{tab:rq1-diversity}
\begin{tabular}{lrrrrrrr}
\toprule
 & \multicolumn{2}{c}{Validity (Plan)} & \multicolumn{2}{c}{Validity (Runtime)} & \multicolumn{3}{c}{Diversity} \\
\cmidrule(lr){2-3}\cmidrule(lr){4-5}\cmidrule(lr){6-8}
Tool & ANSI on & off & ANSI on & off & \#Ops & \#Exprs & \#Trees \\
\midrule
{\tool} & 91.8\% & 99.9\% & 84.4\% & 99.8\% & 25 & 222 & 6{,}975 \\
CometFuzz & 63.5\% & 63.8\% & 53.3\% & 62.4\% & 7 & 130 & 7 \\
SQLancer & 4.7\% & 5.1\% & 4.5\% & 5.1\% & 5 & 29 & 36 \\
\bottomrule
\end{tabular}
\end{table}

\subsection{RQ3: Coverage of Generated Workloads}
\label{sec:rq-coverage}

\begin{figure*}[t]
\centering
\begin{tikzpicture}
\begin{groupplot}[
    group style={group name=coveragegrowth, group size=3 by 1, horizontal sep=1.1cm},
    width=0.36\linewidth, height=4.2cm,
    xmin=0, xmax=60.545, xtick={0,20,40,60},
    xlabel={Testing time (min)},
    title style={font=\bfseries\small},
    tick label style={font=\scriptsize},
    label style={font=\small},
    scaled y ticks=false, scaled x ticks=false,
    y tick label style={/pgf/number format/.cd, fixed, 1000 sep={\mathord{,}}},
    x tick label style={/pgf/number format/.cd, fixed, 1000 sep={\mathord{,}}},
    grid=major, grid style={gray!25},
    legend style={font=\small, draw=none, /tikz/every even column/.append style={column sep=0.5cm}},
    table/col sep=comma,
]
\nextgroupplot[title={Line coverage},
    legend to name=coveragelegend, legend columns=3]
\addplot[color=blue!80!black, very thick] table[x=minutes, y=line] {Figure/data/coverage-discprop.csv};
\addlegendentry{{\tool}}
\addplot[color=red!75!black, very thick, dashed] table[x=minutes, y=line] {Figure/data/coverage-comet.csv};
\addlegendentry{CometFuzz}
\addplot[color=green!50!black, very thick, dash dot] table[x=minutes, y=line] {Figure/data/coverage-sqlancer.csv};
\addlegendentry{SQLancer}
\nextgroupplot[title={Branch coverage}]
\addplot[color=blue!80!black, very thick] table[x=minutes, y=branch] {Figure/data/coverage-discprop.csv};
\addplot[color=red!75!black, very thick, dashed] table[x=minutes, y=branch] {Figure/data/coverage-comet.csv};
\addplot[color=green!50!black, very thick, dash dot] table[x=minutes, y=branch] {Figure/data/coverage-sqlancer.csv};
\nextgroupplot[title={Method coverage}]
\addplot[color=blue!80!black, very thick] table[x=minutes, y=method] {Figure/data/coverage-discprop.csv};
\addplot[color=red!75!black, very thick, dashed] table[x=minutes, y=method] {Figure/data/coverage-comet.csv};
\addplot[color=green!50!black, very thick, dash dot] table[x=minutes, y=method] {Figure/data/coverage-sqlancer.csv};
\end{groupplot}
\node at ($(coveragegrowth c2r1.south) + (0,-1.3cm)$) {\pgfplotslegendfromname{coveragelegend}};
\end{tikzpicture}
\vspace{-1em}
\caption{{\tool} improves code coverage on the optimizer and the runtime executor, compared to existing testing tools. Line, branch, and method coverage of Spark's \texttt{catalyst} and \texttt{execution} packages over one hour of testing on Spark 4.2.0, for {\tool}, CometFuzz~\cite{cometfuzz_github}, and SQLancer's Spark provider~\cite{sqlancer_repo,sqlancer_spark_pr}. Testing time counts each tool's own generation and execution.}
\label{fig:coverage-growth}
\end{figure*}

\paragraph{Setup.}
We compare {\tool} with two Spark SQL testing tools: CometFuzz~\cite{cometfuzz_github}, which generates queries from a grammar and compares Spark with and without the Comet accelerator~\cite{datafusion_comet}, and SQLancer~\cite{sqlancer_repo} with its Spark provider~\cite{sqlancer_spark_pr}. Their oracles target other kinds of faults (\S\ref{sec:rq-new-bugs}), so we compare only the engine code that their generated queries reach, and only on Spark, because neither supports Flink. Each tool runs on Spark 4.2.0 with the default configuration (ANSI mode on), a 30-second timeout per program, four dedicated cores, and its own default data and program size. CometFuzz queries three tables of 1{,}000 rows. SQLancer creates one or two tables, fills them with up to 30 \texttt{INSERT} statements of a few rows each, and generates expressions up to depth 3. Each query of either tool is a single SQL statement. {\tool} runs base workloads without property transformations, each with 1 to 9 unary and up to 3 binary DataFrame operators over sources of 5 columns and 3 to 10 rows, where 5\% of the sources have up to 100 rows instead. Each tool runs for one hour of testing time, which counts only the generation and execution of programs~\cite{klees2018evaluating}. We measure the coverage of Spark's Catalyst and execution packages (\texttt{org.apache.spark.sql.catalyst} and \texttt{org.apache.spark.sql.execution}) with JaCoCo~\cite{jacoco}.

\paragraph{Code coverage.}
Figure~\ref{fig:coverage-growth} plots coverage over testing time. After one hour, {\tool} covers 27{,}298 branches, 23{,}733 methods, and 36{,}882 lines, compared with 14{,}048, 13{,}939, and 25{,}090 for CometFuzz and 12{,}948, 12{,}132, and 21{,}695 for SQLancer. This is 1.9$\times$ the branch coverage of CometFuzz and 2.1$\times$ that of SQLancer. All three tools reach most of their final coverage within the first ten minutes, so a longer budget is unlikely to change the comparison. {\tool} thus exercises more of the engine than both tools, although every program it generates must be valid and property-checkable.

\paragraph{Why does {\tool} reach more of the engine?}
Table~\ref{tab:rq1-diversity} reports two auxiliary measures over 10{,}000 programs per tool. First, most queries of the other two tools never reach the optimizer. With ANSI mode off, Spark plans 99.9\% of the {\tool} programs (the other 0.1\% hit a known optimizer bug) but only 63.8\% of the CometFuzz queries and 5.1\% of the SQLancer queries, which fail almost entirely with analysis errors, mostly type errors such as a comparison of a date with a decimal. ANSI mode, the default since Spark 4.0~\cite{spark_ansi}, raises an error on arithmetic overflow or an invalid cast where Spark would otherwise return \texttt{NULL} or wrap around. {\tool}'s data generator injects boundary values on purpose, such as the largest integer or a string that is not a number, and does not model which values each expression accepts, because that would make generation expensive. Under ANSI mode, about 8\% of its programs therefore fail during planning, where Spark already evaluates expressions over in-memory inputs, and about 7\% more fail at run time. Second, {\tool}'s plans are more varied. To compare plan shapes, we anonymize each physical plan by removing its expressions and column names, which leaves only the tree of physical operators. {\tool}'s plans contain 25 distinct physical operators, 222 distinct expression APIs, and 6{,}975 distinct operator trees, compared with 7, 130, and 7 for CometFuzz and 5, 29, and 36 for SQLancer. Half of {\tool}'s plans (51\%) combine a join with an aggregation, which no plan of the other two tools does, and bugs such as Flink-1 (an aggregate above a join) need such compositions. CometFuzz covers 16 expressions that {\tool} does not, mostly trigonometric, bitwise, and hexadecimal functions.

\paragraph{Cost.}
In one hour, {\tool} submitted 7{,}100 programs, CometFuzz 8{,}586 queries, and SQLancer 13{,}000 TLP attempts, that is, 0.51, 0.42, and 0.28 seconds per program under coverage instrumentation. A {\tool} program takes longer because it is a multi-stage DataFrame workload over several sources rather than a single query. Almost all of its cost lies in Spark. Without instrumentation, in one long-lived SparkSession warmed up with 10 workloads, a base workload takes 193\,ms on average (median 178\,ms) over 1{,}000 executable workloads: 2.0\,ms for generation, including the analysis that tracks schemas and types, 32\,ms for constructing the DataFrames in Spark and analyzing their schemas, and 158\,ms for planning and execution. A property test additionally executes the transformed workload and compares the results. In the 12-hour Spark campaign of RQ1, six workers ran 498{,}333 property tests, about 6{,}900 per worker and hour.

\begin{figure*}[ht]
\centering
\begin{tikzpicture}
\begin{groupplot}[
    group style={
        group name=scalesweepvalidity,
        group size=4 by 1,
        horizontal sep=0.4cm,
    },
    scale only axis,
    width=0.2\linewidth,
    enlarge x limits=0.06,
    height=2.5cm,
    title style={font=\bfseries\scriptsize},
    tick label style={font=\scriptsize},
    label style={font=\scriptsize},
    scaled y ticks=false,
    y tick label style={font=\scriptsize},
    legend style={font=\footnotesize, draw=none, fill=none, /tikz/every even column/.append style={column sep=0.4cm}},
]

\nextgroupplot[
    title={Schema width},
    ymin=0, ymax=108,
    ytick={0,50,100},
    ylabel={Rate (\%)},
    xmode=log,
    log basis x=2,
    xtick={5,10,20,40,80},
    xticklabels={5,10,20,40,80},
    xlabel={\# columns},
    legend to name=sweepvaliditylegend,
    legend columns=3,
]
\addplot[color=green!50!black, thick, dashed, mark=square*, mark size=1.1pt] coordinates {
(5,100.0) (10,100.0) (20,100.0) (40,100.0) (80,100.0)
};
\addlegendentry{Plan-validity rate}
\addplot[color=red!75!black, thick, dotted, mark=triangle*, mark size=1.3pt] coordinates {
(5,3.85) (10,1.96) (20,3.85) (40,1.96) (80,1.96)
};
\addlegendentry{ANSI data-error rate}

\nextgroupplot[
    title={Rows},
    ymin=0, ymax=108,
    ytick={0,50,100},
    yticklabels={,,},
    xmode=log,
    xtick={10,100,1000,10000,50000},
    xticklabels={10,100,1k,10k,50k},
    xlabel={\# rows},
]
\addplot[color=green!50!black, thick, dashed, mark=square*, mark size=1.1pt] coordinates {
(10,100.0) (100,100.0) (1000,100.0) (10000,100.0) (50000,100.0)
};
\addplot[color=red!75!black, thick, dotted, mark=triangle*, mark size=1.3pt] coordinates {
(10,1.96) (100,1.96) (1000,1.96) (10000,1.96) (50000,7.41)
};

\nextgroupplot[
    title={Binary-operator ratio},
    ymin=0, ymax=108,
    ytick={0,50,100},
    yticklabels={,,},
    xtick={0.15,0.3,0.5,0.7,0.85},
    xlabel={$r_{\mathrm{bin}}$},
]
\addplot[color=green!50!black, thick, dashed, mark=square*, mark size=1.1pt] coordinates {
(0.15,100.0) (0.3,100.0) (0.5,100.0) (0.7,100.0) (0.85,100.0)
};
\addplot[color=red!75!black, thick, dotted, mark=triangle*, mark size=1.3pt] coordinates {
(0.15,7.27) (0.3,3.85) (0.5,1.96) (0.7,1.96) (0.85,1.92)
};

\nextgroupplot[
    title={Depth},
    ymin=0, ymax=108,
    ytick={0,50,100},
    yticklabels={,,},
    xmode=log,
    log basis x=2,
    xtick={3,6,12,24,48,96},
    xticklabels={3,6,12,24,48,96},
    xlabel={Workload DAG depth},
]
\addplot[color=green!50!black, thick, dashed, mark=square*, mark size=1.1pt] coordinates {
(3,100.0) (6,100.0) (12,100.0) (24,100.0) (48,100.0) (96,100.0)
};
\addplot[color=red!75!black, thick, dotted, mark=triangle*, mark size=1.3pt] coordinates {
(3,1.96) (6,5.66) (12,21.88) (24,32.43) (48,40.48) (96,59.38)
};

\end{groupplot}

\node at ($(scalesweepvalidity c1r1.south)!0.5!(scalesweepvalidity c4r1.south) + (0,-1.25cm)$)
    {\pgfplotslegendfromname{sweepvaliditylegend}};

\end{tikzpicture}
\vspace{-1em}
\caption{
Generated workloads stay well-formed as data and computation grow:
plan validity is 100\% at every point as columns, rows, binary-operator ratio
$r_{\mathrm{bin}}$, and DAG depth each scale from the defaults (5 columns,
10 rows, $r_{\mathrm{bin}}=0.15$, depth 3) on Spark 4.2.0. ANSI data errors
come from injected boundary values and vanish with ANSI mode off.}
\label{fig:scale-sweep-validity}
\end{figure*}

\subsection{RQ4: Validity and Scalability}
\label{sec:rq-scalability}

RQ4 tests whether the base generator, without property checks, stays valid as workloads grow, on Spark 4.2.0 with ANSI mode on. Starting from defaults of 5 columns, 10 rows, binary-operator ratio $r_{\mathrm{bin}}=0.15$ (the probability that a workload node is a binary operator such as a join or a union), and depth 3, we vary one dimension at a time: up to 80 columns, 50{,}000 rows, $r_{\mathrm{bin}}=0.85$, and depth 96. For each configuration, we sample attempts until 50 programs complete, with a 60-second timeout per program. We report the validity rate (programs that Spark accepts without an analysis error), the rate of ANSI data errors (programs that fail at run time because a generated value overflows or cannot be cast), and the average execution time. Validity here means that the generator builds well-formed workloads, not that every property holds on them (\S\ref{sec:rq-new-bugs}).

\paragraph{Results.}
Spark accepts every generated program at every configuration without an analysis error (Figure~\ref{fig:scale-sweep-validity}). At large depths, the generator sometimes runs out of the column types that an operator needs, such as a grouping key, and retries. At depth 96, a third of the attempts are retried, and each costs only milliseconds of generation. ANSI data errors stay at or below 9\% on every axis except depth, where they grow from 2\% at depth 3 to 59\% at depth 96, because each additional operator is another chance for a generated boundary value to overflow or to fail a cast. With ANSI mode off, the same sweep has no data errors and the same 100\% program validity. Besides data errors, 3 of the 1{,}230 generated programs failed inside Spark. Two of them hit a known optimizer bug (\texttt{PushProjectionThroughUnion}), and one raised an internal error at depth 96. Average execution time stays below 1.6\,s at every configuration except 50{,}000 rows (2.7\,s).

\section{Limitations and Threats to Validity}
\label{sec:limitations}

\paragraph{Correctness of bindings.}
\tool{} checks each binding against the signature of its meta-property (Section~\ref{sec:mp-typing}), which guarantees schema-valid tests but not a semantically correct binding. We chose this trade-off on purpose: a formal proof would need a full model of the engine's semantics, whereas a binding is short and easy to test in isolation, and a mistake surfaces as a violation that we triage once (Section~\ref{sec:mp-laws}). In our campaigns, false alarms stayed below one per 10{,}000 tests. Similarly, the generator does not model which values each expression accepts. This keeps generation cheap and lets it inject the boundary values that trigger many bugs, at the cost of some programs that fail with data errors under ANSI mode and are discarded (Sections~\ref{sec:rq-coverage} and~\ref{sec:rq-scalability}).

\paragraph{Scope.}
\tool{} targets deterministic wrong results of DataFrame workloads over in-memory data, the class of bugs that is hardest to detect because the engine reports no error. Bugs in data sources, catalogs, streaming, other front ends, or multi-executor execution lie outside this scope. Within it, the seven meta-properties express 55 of the 65 testable Spark correctness issues, and seven of the remaining ten point to two further meta-properties (Section~\ref{sec:rq-generality}). Adding them needs no change to the generator. Like any testing tool, \tool{} does not guarantee that it exposes a covered bug, because it must generate the triggering workload and data itself. The same freedom lets it find bugs through workloads that their reporters never wrote, and richer operator combinations, values, and input sizes in the generator would benefit all meta-properties at once.

\paragraph{Threats to validity.}
The RQ2 classification involves judgment. Claude Opus 5.5 classified the issues and wrote the bindings, and an author reviewed the classification and the coverage decisions. To keep the judgment checkable, a bug counts as covered only if its concrete property fails on a buggy release and holds on the fixed one, and the end-to-end campaigns use the unmodified generator of RQ1. RQ2 studies Spark issues with the \emph{correctness} label, a set that the Spark developers curate independently of our work. We evaluate \tool{} on two widely used engines with different optimizers, Spark and Flink. Because the meta-properties build on engine-independent principles, we expect them to carry over to other DISC engines, which would need new bindings and engine-specific support code (Section~\ref{sec:portability}). RQ3 compares code coverage on Spark only, because neither baseline supports Flink, and coverage measures how much of the engine a tool reaches rather than how many bugs its oracle detects. We therefore report the bugs separately in RQ1.

\section{Related Work}

\paragraph{Property-based testing and where properties come from.}
QuickCheck~\cite{claessen2000quickcheck} popularized PBT, which checks
properties over generated inputs and shrinks failing
cases~\cite{scalacheck,hypothesis,jqwik,hughes2016experiences}. It takes the
properties as given: each is written by hand, one at a time. PBT tools for
data-intensive systems inherit this assumption. sscheck~\cite{riesco2018sscheck}
and FlinkCheck~\cite{espinosa2019flinkcheck} check user-written bounded
temporal-logic properties of streaming applications, and
DiffStream~\cite{kallas2020diffstream} checks user-provided dependence relations
between stream outputs. In each, the developer writes every property for one
application. Prior work addresses where properties come from in two ways.
QuickSpec~\cite{quickspec} discovers equational laws from a
library's functions, whereas \tool{} starts from known laws and reuses them.
Law-checking tools~\cite{quickcheckclasses,testingtypeclass} reuse a law
across instances: a type-class law is written once and checked for every
instance, much as a meta-property is written once and checked for every
binding. The two differ in what they quantify over and what they test. A
type-class law quantifies over values and tests the instance, whereas a meta-property
quantifies over program contexts, which \tool{} generates around its holes,
and tests the engine that evaluates them. 

\paragraph{Testing Data-Intensive Applications and Systems.}
Prior work on testing data-intensive applications focuses primarily on
test-input generation and coverage improvement for user applications rather
than the underlying frameworks. BigFuzz~\cite{zhang2021bigfuzz} abstracts
dataflow operators, DepFuzz~\cite{depfuzz_fse2023} uses dependency-aware
mutation, BigTest~\cite{gulzar2019white} targets path-aware symbolic execution
testing, and NaturalSym~\cite{naturalsym_fse2024} optimizes for input
naturalness. While effective for exploring execution paths within user
applications, these techniques rely on weak oracles, such as crashes,
timeouts, or unhandled exceptions, and cannot verify semantic correctness in
the underlying frameworks. Differential testing tools such as
SparkFuzz~\cite{ghit2020sparkfuzz} and CometFuzz~\cite{cometfuzz_github} obtain
a strong oracle from a second reference implementation or a second configuration.
\tool{} instead instantiates strong oracles by binding meta-properties.

\paragraph{Metamorphic testing.}
Metamorphic testing checks a relation between the outputs of related inputs.
Database testing has made it practical: TLP~\cite{rigger2020finding_tlp}
partitions a query by a predicate, NoREC~\cite{norec_fse2020} rewrites it to
defeat the optimizer, and CODDTest~\cite{coddtest_sigmod2025} rewrites constant
subqueries. PQS~\cite{pqs_osdi2020} instead checks that a constructed pivot row
is retrieved. Related oracles target Datalog~\cite{zhang2024finding}, graph
systems~\cite{gslicer_sigmod2025}, and stream processing
engines~\cite{kroner2025achilles}. Most of these relations are fixed
perturbations: the tool transforms an input in one hard-coded way and compares
the outputs for equality. Metamorphic relation
patterns~\cite{segura2019mrpatterns} describe recurring relations between
query outputs, such as equivalence, subset, and disjointness, as informal
templates that guide engineers in identifying relations. \tool{}'s
meta-properties make such templates executable and reusable. Their holes are
typed and may be higher-order, ranging over expressions, lambdas,
user-defined functions, operators, and the oracle relation itself, so a
binding is checked against a signature, and every test generated from a
type-correct binding is schema-valid (Section~\ref{sec:mp-typing}).
Section~\ref{sec:mp-laws} discusses when its oracle holds. Relations need not be equalities,
as with the bound for \texttt{avg}. Several existing oracles are close to concrete instances of meta-properties:
TLP resembles P5, NoREC resembles P3, and P5 and P7 can be read as typed,
executable forms of relation patterns.

\section{Conclusion}
Property-based testing catches the silent wrong-result bugs that
crash-detecting tools miss, but each property must come from somewhere, and writing them one
API at a time cannot keep pace with frameworks whose APIs keep growing. For
data-intensive computing frameworks, this paper answers where properties come
from: textbook principles that the frameworks already obey. \tool{} captures each
principle as a \emph{meta-property}, a property with holes that a developer fills
with a short binding. Each meta-property has a typing rule, so a binding is checked once and every test generated from it is schema-valid.

Seven meta-properties yield 207 concrete properties, with bindings averaging
under 20 lines. With them, \tool{} found 17 previously unknown wrong-result
bugs in the latest releases of Spark and Flink, nine of which need relations
that existing SQL testing oracles such as TLP and NoREC do not state, and its meta-properties express 55 of 65
testable in-scope historical Spark correctness issues. Its generator, shared
by both engines, produces workloads that Spark accepts without an analysis
error as schemas, data, and workload depth grow, and that cover 1.9$\times$
the branches of CometFuzz.

More broadly, a principle written once as a meta-property can be reused many times:
across APIs, workloads, data, and engine versions. We expect this
\emph{write once, reuse many times} recipe to apply to other domains with rich
algebraic structure, such as tensor computation frameworks.

\bibliographystyle{IEEEtran}
\bibliography{reference}

\appendix
\section{Typing Rules of All Meta-Properties}
\label{app:rules}
Figure~\ref{fig:mp-rules-all} gives the typing rules of P2 to P7, in addition to the rule of P1 in Figure~\ref{M-fig:mp-rules}. In P5, $R_{\uplus}(O, O_1, O_2, O_3) \Leftrightarrow O \eqbag O_1 \uplus O_2 \uplus O_3$. In P3, $\lceil\cdot\rceil_k$ adapts each downstream operator to carry the tag column $k$ (Appendix~\ref{app:p3}).

\begin{figure}[h]
\centering
\small
\mprset{sep=1.2em}
\begin{mathpar}
\inferrule*[right=\rn{P2}]{
  \bnd{f} \text{ declared } \bar\tau \to \tau \\ \bar x{:}\bar\tau \vdash_{\mathrm{row}} \bnd{e} : \tau \\
  \vdash C : (\Sigma_h \vdash_{\mathrm{row}} \tau) \Rightarrow (\vS \Rightarrow \Sigma) \\ \Sigma_h \vdash \sigma : \bar x{:}\bar\tau
}{\vdash \tst{C[\mathsf{udf}_{\tau}(f)(\bar x)\sigma],\ C[e\sigma]}{\eqbag} : \vS}
\and
\inferrule*[right=\rn{P3}]{
  \vdash \Qpre : \vS \Rightarrow \Sigma_1 \\ \Sigma_1 \vdash_{\mathrm{row}} p : \mathsf{bool} \\
  \vdash \Qpost : \Sigma_1 \Rightarrow \Sigma \\ \vdash \lceil \Qpost \rceil_k : \Sigma_1, k{:}\mathsf{bool} \Rightarrow \Sigma, k{:}\mathsf{bool}
}{\vdash \tst{[\,\Qpre;\ \mathsf{filter}(p);\ \Qpost\,],\ [\,\Qpre;\ \mathsf{withColumn}(k, p);\ \lceil\Qpost\rceil_k;\ \mathsf{filter}(k);\ \mathsf{drop}(k)\,]}{\eqbag} : \vS}
\and
\inferrule*[right=\rn{P4}]{
  \vdash Q : \vS \Rightarrow \Sigma \\ \Sigma \vdash_{\mathrm{row}} c : \tau \\ \Sigma \vdash_{\mathrm{row}} k : \tau_k \\\\
  \bnd{f} : \Bag{\tau} \to \tau' \\ \bnd{R} \subseteq \tau' \times \Bag{\tau_k \times \tau'}
}{\vdash \tst{[\,Q;\ \mathsf{agg}(f(c))\,],\ [\,Q;\ \mathsf{groupBy}(k).\mathsf{agg}(f(c))\,]}{R} : \vS}
\and
\inferrule*[right=\rn{P5}]{
  \vdash Q : \vS \Rightarrow \Sigma \\ \Sigma \vdash_{\mathrm{row}} p : \mathsf{bool}
}{\vdash \tst{Q,\ [\,Q;\ \mathsf{filter}(p)\,],\ [\,Q;\ \mathsf{filter}(\neg p)\,],\ [\,Q;\ \mathsf{filter}(p\ \mathsf{is\ null})\,]}{R_{\uplus}} : \vS}
\and
\inferrule*[right=\rn{P6}]{
  \vdash \Qpre : \vS \Rightarrow \Sigma_1 \\ \vdash \bnd{M} : \Sigma_1 \Rightarrow \Sigma_1 \\ \vdash \Qpost : \Sigma_1 \Rightarrow \Sigma
}{\vdash \tst{[\,\Qpre;\ \Qpost\,],\ [\,\Qpre;\ M;\ \Qpost\,]}{\eqbag} : \vS}
\and
\inferrule*[right=\rn{P7}]{
  \vdash Q : \vS \Rightarrow \Sigma \\ \vdash \bnd{\mathrm{op}} : \Sigma \Rightarrow \Sigma' \\ \bnd{R_{\mathrm{op}}} \subseteq \Bag{\Sigma} \times \Bag{\Sigma'}
}{\vdash \tst{Q,\ [\,Q;\ \mathrm{op}\,]}{R_{\mathrm{op}}} : \vS}
\end{mathpar}

\caption{Typing rules of P2 to P7. Highlighted premises come from the binding.}
\label{fig:mp-rules-all}
\end{figure}

\section{\texorpdfstring{Downstream Adaptation in P3}{Downstream Adaptation in P3}}
\label{app:p3}
P3 compares a workload that filters by a predicate $p$ before a downstream fragment $Q_{\mathrm{post}}$ with one that stores $p$ in a tag column $k$, runs the adapted fragment $\lceil Q_{\mathrm{post}}\rceil_k$, and keeps the rows where $k$ is true at the end. Table~\ref{tab:p3-adapt} lists how \tool{} adapts each kind of downstream operator and why the two sides then agree. $Q_{\mathrm{post}}$ is a chain of operators, and its other inputs, such as the right side of a join, do not carry $k$. A row is \emph{kept} if $p$ is true on it and \emph{removed} otherwise, including when $p$ is null.

\begin{table}[h]
\caption{Downstream operators in P3, how \tool{} adapts each one to carry the tag column $k$, and why the adapted operator agrees with filtering by $p$ first.}
\label{tab:p3-adapt}
\centering
\small
\setlength{\tabcolsep}{4pt}
\renewcommand{\arraystretch}{1.2}
\begin{tabularx}{\textwidth}{@{}>{\raggedright\arraybackslash}p{4.2cm} >{\raggedright\arraybackslash}p{3.9cm} >{\raggedright\arraybackslash}X@{}}
\toprule
\textbf{Downstream operator} & \textbf{Adaptation $\lceil\mathrm{op}\rceil_k$} & \textbf{Why it agrees} \\
\midrule
Row-local: \texttt{filter} by a predicate that does not read $k$, \texttt{withColumn}, rename, \texttt{drop} or \texttt{select} of other columns, \texttt{fillna}, \texttt{replace}, and \texttt{dropna} on a column subset, \texttt{explode}, \texttt{unpivot} with $k$ as an identifier, \texttt{orderBy}, \texttt{repartition}, \texttt{coalesce}
 & unchanged, and $k$ is copied to every output row
 & each output row depends only on its input row \\
\addlinespace
Joins with the tagged input on the left: inner, left outer, left semi, left anti, cross, lateral
 & unchanged, and $k$ comes from the left row
 & the output rows of a left row depend only on that row and the right input \\
\addlinespace
\texttt{distinct}, \texttt{dropDuplicates} over all columns
 & deduplicate over the row together with $k$
 & a row that is kept at least once survives exactly once after the final filter \\
\addlinespace
\texttt{union}, \texttt{intersect}, \texttt{intersectAll}, \texttt{exceptAll}, \texttt{subtract}
 & the other input gets $k = \mathsf{true}$, and the tagged input is on the left of \texttt{exceptAll} and \texttt{subtract}
 & a kept row behaves as on the filtered side, and a removed row matches no row of the other input and is dropped at the end \\
\addlinespace
Final aggregation $\mathsf{groupBy}(g).\mathsf{agg}(h(x))$ with an $h$ that ignores nulls, or \texttt{count(*)}
 & $\mathsf{groupBy}(g)$\newline$.\mathsf{agg}(h(\mathsf{when}(k, x)))$, with \texttt{count(*)} replaced by $\mathsf{count\_if}(k)$, and the output tag $k := \mathsf{count\_if}(k) > 0$, or $k := \mathsf{true}$ for a global aggregation
 & $h$ sees exactly the kept rows of each group, and the final filter keeps exactly the groups with a kept row \\
\midrule
Excluded: right and full outer joins
 & --
 & they pad unmatched right rows, and which rows are unmatched depends on which left rows exist \\
\bottomrule
\end{tabularx}
\end{table}

\end{document}